**Tuning ultrasmall theranostic nanoparticles for MRI contrast and radiation dose amplification**

*Needa Brown[#\*], Paul Rocchi[#], Léna Carmès, Romy Guthier, Meghna Iyer, Léa Seban, Toby Morris, Stephanie Bennett, Michael Lavelle, Johany Penailillo, Ruben Carrasco, Chris Williams, Elizabeth Huynh, Zhaohui Han, Evangelia Kaza, Tristan Doussineau, Sneh M. Toprani, Xingping Qin, Zachary D. Nagel, Kristopher A. Sarosiek, Agnès Hagège, Sandrine Dufort, Guillaume Bort, François Lux, Olivier Tillement, and Ross Berbeco[\*]*

[#]Co-first authors

[\*]Corresponding authors

Needa Brown, Meghna Iyer

Department of Physics, Northeastern University, Boston 02115, USA

Email: ne.brown@northeastern.edu

Needa Brown, Romy Guthier, Meghna Iyer, Léa Seban, Toby Morris, Stephanie Bennett, Michael Lavelle, Chris Williams, Elizabeth Huynh, Zhaohui Han, Evangelia Kaza, Ross Berbeco

Department of Radiation Oncology, Brigham and Women's Hospital, Dana-Farber Cancer Institute, and Harvard Medical School, Boston 02115, USA

Email: ross_berbeco@dfci.harvard.edu

Paul Rocchi, Léna Carmès, Tristan Doussineau, Sandrine Dufort

NHTherAguix, Lyon 69100, France

Paul Rocchi, Léna Carmès, Guillaume Bort, François Lux, Olivier Tillement

Institut Lumière-Matière, UMR 5306, Université Lyon1-CNRS, Université de Lyon, Villeurbanne Cedex 69100, France

Romy Guthier, Toby Morris, Michael Lavelle

Department of Physics and Applied Physics, University of Massachusetts Lowell, Lowell 01854, USA

Johany Penailillo, Ruben Carrasco

Department of Pathology, Harvard Medical School and Dana-Farber Cancer Institute, Boston 02115, USA

Sneh M. Toprani, Xingping Qin, Zachary D. Nagel, Kristopher A. Sarosiek

John B. Little Center for Radiation Sciences, Department of Environmental Health, Harvard T.H. Chan School of Public Health, Boston, MA, 02115, USA

Xingping Qin, Kristopher A. Sarosiek

Laboratory of Systems Pharmacology, Harvard Program in Therapeutic Science, Department of Systems Biology, Harvard Medical School, Boston, MA, 02115, USA

Molecular and Integrative Physiological Sciences Program, Harvard T.H. Chan School of Public Health, Boston, MA, 02115, USA

Department of Medical Oncology, Dana-Farber Cancer Institute/ Harvard Cancer Center, Boston, MA, 02115, USA

Agnès Hagège

Institut des Sciences Analytiques, Université de Lyon, CNRS, Université Claude Bernard Lyon 1, UMR 5280, 69100, Villeurbanne, France

François Lux

Institut Universitaire de France (IUF), Paris 75005, France

**Abstract**

The introduction of magnetic resonance (MR)-guided radiation treatment planning has opened a new space for theranostic nanoparticles to reduce acute toxicity while improving local control. In this work, second-generation AGuIX® nanoparticles (AGuIX-Bi) are synthesized and validated to maintain MR positive contrast while further amplifying the radiation dose by replacing some $Gd^{3+}$ cations with higher Z $Bi^{3+}$. Next-generation nanoparticles are based on the AGuIX® platform, which is currently being evaluated in multiple Phase II clinical trials in combination with radiotherapy. In this clinically scalable methodology, AGuIX® is used as an initial chelation platform to exchange $Gd^{3+}$ for $Bi^{3+}$. AGuIX-Bi nanoparticles are synthesized with three ratios of Gd/Bi, each maintaining MR contrast while further amplifying radiation dose relative to $Bi^{3+}$. In a non-small cell lung cancer model, increased $Bi^{3+}$ is associated with more DNA damage and improves *in vivo* efficacy with a statistically significant delay in tumor growth and 33% complete regression. The addition of $Bi^{3+}$ by our synthetic method leads to nanoparticles that present slightly altered pharmacokinetics with lengthening of the period of high tumor accumulation with no observed evidence of toxicity. We demonstrate the safety and enhanced efficacy of AGuIX-Bi at the selected ratio of 30Gd/70Bi, both crucial steps toward patient translation.

**Introduction**

The first gadolinium (Gd)-based contrast agent, Magnevist®, entered the global clinics in 1988 [15]. Since then, advances in contrast agents and Magnetic Resonance Imaging (MRI) capabilities have allowed several novel formulations to enter the clinic including the theranostic nanoparticle (NP) AGuIX® (Activation and Guidance of Irradiation by X-Ray). AGuIX® NPs were developed by NH TherAguix and first described in 2011 [16]. They are sub-5 nm NPs composed of polysiloxane and a high concentration of Gd (~15 Gd per NP) for MR guidance and local radiation dose amplification. Gadolinium chelates within AGuIX® serve a dual role as a MR-contrast agent and a local amplifier, emitting more low-energy photoelectrons and Auger electrons during radiation therapy. The small size enables high tumor-specific accumulation, taking advantage of the enhanced permeation and retention (EPR) effect and quick renal clearance. AGuIX® has shown promise in phase I dose escalation studies on 15 patients with continuation into multiple phase II trials (NCT03818386 & NCT04789486) for the treatment of brain metastases, lung cancer, and pancreatic cancer in combination with radiation therapy [14, 17]. AGuIX® is one of only two radiosensitizer nanoparticles to make it into clinical trials, the other being hafnium oxide ($Z_{Hf}$ = 72) nanoparticles, NBTXR3, developed by Nanobiotix. Compared to AGuIX®, NBTXR3 relies on intratumoral injections thus limiting therapies to only superficially accessible tumors and providing limited diagnostic capabilities [18].

Although AGuIX® NPs have demonstrated promising theranostic potential, treatments do not lead to complete regression. A push to improve the therapeutic potential of conventional AGuIX® NPs while still maintaining MR contrast motivated the formulation of second-generation bismuth (Bi)-gadolinium nanoparticles (AGuIX-Bi). In this proof-of-concept study, we specifically chose to design AGuIX-Bi NPs to address the unique imaging and therapeutic challenges associated with centrally-located non-small cell lung cancer (NSCLC) in alignment with objectives of the ongoing

Nano-SMART clinical trial (NCT04789486). Being the third most common type of carcinoma, lung tumors account for approximately 25% of cancer related deaths and are estimated to lead to over 238,340 new cases in 2023 alone with 85% consisting of NSCLCs [1]. A unique challenge for NSCLC radiation therapy is the constant motion associated with the cardiac and respiratory cycles within the lungs [9]. This motion can lead to inaccurate tumor dose delivery and increased risk for surrounding healthy tissue damage, especially in centrally-located lesions with high rates of proximal critical organ (central airways, esophagus, and vascular structures) toxicity. Stereotactic body radiation therapy (SBRT), which involves a high dose per fraction (>6 Gy fraction$^{-1}$), is a proven method for the treatment of early-stage NSCLC patients, achieving local control rates of 90% [19, 20]; however, dose-limiting toxicities are a major obstacle restricting this treatment option for thousands of patients each year. Breath-holds are often used to mitigate motion artifacts and proximal toxicities; however, NSCLC patients often have other comorbidities that make breath-holds difficult. Alternatively, image guidance via MRI is being integrated into treatment workflows, improving tumor dose delivery while reducing patient burden [9]. Although MRI provides excellent soft tissue contrast at most body sites, the unique high air content and low tissue density within the lungs dramatically reduces the signal-to-noise ratio. This coupled with motion due to the cardiac and respiratory cycles leads to susceptibility artifacts and image degradation [9]. Recent literature with tumor specific gadolinium-based nanoparticles [14] has shown potential for improved contrast and signal-to-noise ratios when coupled with MRI. The introduction of MR-guided radiation therapy within the clinic has opened up a new space for theranostic nanoparticles like AGuIX® that will enable better tumor delineation, local radiation amplification, and sparing of healthy tissue.

In this paper, second-generation NPs have been formulated which maintain MR positive contrast while further amplifying the radiation dose effect with additional chelation of bismuth.

Compared to AGuIX®, these AGuIX-Bi NPs incorporate higher atomic number (Z) $Bi^{3+}$ ($Z_{Bi}$ = 83) cations in addition to $Gd^{3+}$ ($Z_{Gd}$ = 64) enhancing their theranostic applications. The probability of the photoelectric effect increases proportional to the atomic number of the atom ($Z^3$) and decreases when the energy of the incident photon increases ($E^3$). In soft tissue, the photoelectric effect has a relatively small contribution at energies >30 keV, however, it is the dominant interaction for high-Z nanoparticles leading to the deposition of localized focal ionizing radiation. The secondary photoelectrons and Auger electrons have a short range, a few nanometers to micrometers, leading to a higher local radiation dose effect in tissues with NP uptake [21]. High energy (≥6 MV) photon beams are conventionally used in the clinic to increase skin sparing, however, these beams are comprised of a spectrum of energies with substantial amounts of lower energy photons (<150 kV) which can interact with high Z NPs leading to local dose amplification. This is particularly true for modern flattening filter-free (FFF) beams. In addition to the inherent low energy photons within the spectrum, scatter within the tissue as a function of depth, field size, and distance from the central axis as well as removal of the flattening filter can further contribute to increased lower energy photons [22]. Although physical dose amplification by the photoelectric effect is important, in almost all experiments the degree of observed biological dose enhancement of high Z NPs is much greater than the predicted physical dose, highlighting unknown cellular radiobiological responses that are not driven simply by increasing radiation dose [23, 24]. Other chemical and biological contributions are still being investigated including a recent study with AGuIX® suggesting several biological mechanisms including increased cellular DNA damage due to inhibition of homologous recombination repair pathway proteins, increased apoptosis via activation of caspase-3, and regulation of ferroptotic cell death by inhibiting the NRF2-GSH-GPX4 signaling pathway [25].

In this work, three different gadolinium/bismuth ratio nanoparticles have been synthesized respectively 70Gd/30Bi, 50Gd/50Bi, and 30Gd/70Bi (molar ratio). The formulation methodology developed is a stable and clinically scalable process in which AGuIX® is seen as a chelation platform NP initially filled with Gd that is replaced with Bi at varying ratios. Similar to AGuIX®, the ultrasmall size of AGuIX-Bi nanoparticles facilitates EPR-directed tumor-specific accumulation and kidney clearance. The addition of $Bi^{3+}$ maintains similar physio-chemical characteristics to AGuIX®, retains MRI positive contrast due to the remaining $Gd^{3+}$ cations while amplifying DNA damage and therapeutic efficacy, and increases *in vivo* pharmacokinetics thereby lengthening the period of high tumor accumulation. Given the promise of AGuIX® in phase I and ongoing phase II clinical trials, these preliminary studies are necessary to lay the foundation for the translation of these next-generation AGuIX-Bi nanoformulations for MR-guided radiation therapy.

**Results and Discussion**

**AGuIX® as a tunable chelation platform for next-generation nanoparticles**

AGuIX® are ultrasmall nanoparticles composed of a polysiloxane core with covalently grafted gadolinium chelates. DOTAGA (2-(4,7,10-tris(carboxymethyl)-1,4,7,10-tetraazacyclododecan-1-yl) pentanedioic acid), a derivative of DOTA (1,4,7,10-tetraazacyclododecane-1,4,7,10-tetraacetic acid), is a macrocyclic molecule known for strong binding constants of complexes with trivalent cations, e.g. $Bi^{3+}$ ($\log_{10}(K) = 30.3$) [26, 27] or $Gd^{3+}$ ($\log_{10}(K) = 25.6$) [26, 28]. DOTAGA was chosen due to the conservation of the DOTA structure that ensures strong $Gd^{3+}$ chelation even after grafting of the inorganic matrix ($\log_{10}(K) = 24.78$) [29]. The high stability comes from the macrocyclic nature of the ligand and the ability of the carboxylic acid groups to chelate in a deprotonated state [28]. During formulation, gadolinium dissociation from the covalently grafted DOTAGA@($Gd^{3+}$) groups on AGuIX® was performed by protonating the chelate in highly acidic conditions. AGuIX® NPs were placed in a 1 M HCl acidic medium driving competitive chelate protonation leading to the release of some $Gd^{3+}$ cations [28, 30]. The solution was further heated to 50°C during the release step, favoring the dissociation kinetics. Using a reverse phase C4 HPLC column, the free $Gd^{3+}$ was easily separated from the remaining mixture components. As illustrated in **Figure 1**, the released gadolinium was monitored via Inductively coupled plasma-Mass spectrometry (ICP-MS) all along the process. At the end of the first release step, the amount of gadolinium initially released should be directly correlated with the amount of free DOTAGA in the final product. To obtain NPs with 70% of the initial DOTAGA(Gd) being replaced by DOTAGA(Bi), the gadolinium release had to be pushed to at least 70% of the initial Gd amount at the time the reaction was stopped. This Gd release level was reached after 19 h of reaction. At this point, the area of the free Gd signal recovered was 78.5% of the reference area (**Figure 1A**) indicating the end of the release step. To stop the freed $Gd^{3+}$ from re-chelating, purification was performed using the Vivaflow 200 ultrafiltration system (5 kDa cutoff). High-performance liquid chromatography-ICP/MS (HPLC-ICP/MS) confirmed the absence of free $Gd^{3+}$ after purification drying.

Before moving to the bismuth chelation step, a free DOTAGA titration was performed to confirm the real amount of free DOTAGA in the starting material. The result showed a free DOTAGA level of 0.64 µmol $mg^{-1}$ of the product (**Figure S1**) which represented 86% of the initial DOTAGA($Gd^{3+}$) content of the AGuIX® product as can be seen in **Figure 1B**. The bismuth chelation was divided into three steps to reach the three different Bi targets. After each step, the free DOTAGA content was measured (**Figure S2-S4**). We observed that the experimental chelation followed closely with the expected chelation yield moving from 86% free DOTAGA to 56%, from 56% to 34%, and from 34% to 14% (**Figure 1A**). To obtain the 3 final products, each intermediate product was chelated with the right amount of missing $Gd^{3+}$ (**Figure 1C**) as a final step. The resulting products, AGuIX-Bi 70Gd/30Bi, 50Gd/50Bi, and 30Gd/70Bi, have been purified using a Vivaflow 200 ultrafiltration system (5 kDa cutoff) to ensure the absence of remaining free $Gd^{3+}$ and $Bi^{3+}$ cations in solution.

Previous preliminary work demonstrated the efficacy of bismuth incorporation [31], however optimization of Gd/Bi ratios was not investigated. In the Detappe et al. work, the gadolinium-bismuth particles were obtained through a post-functionalization process of the AGuIX® in which DOTA-NHS was used to react with free amino functional groups at the surface of the nanoparticle. The bismuth was then added on the newly grafted DOTA, but the ratio of Bi was limited by the grafting of DOTA-NHS on the polysiloxane matrix and could not exceed 30-40% due to steric hindrance and the lack of amino functions at the surface of the polysiloxane. Moreover, tuning the ratio of Bi required changing the number of DOTA grafted at the surface resulting in changes in surface charge and size which complicates comparisons between NPs with different Gd/Bi ratios. In the current work, AGuIX® was used as a chelation platform NP initially filled with Gd, thus providing a more scalable formulation for clinical translation. Through an acidic treatment, we were able to remove the initial Gd and regenerate free DOTAGA. The newly generated free DOTAGA was used to achieve the desired complexation of different elements at different ratios making this a more tunable and scalable platform.

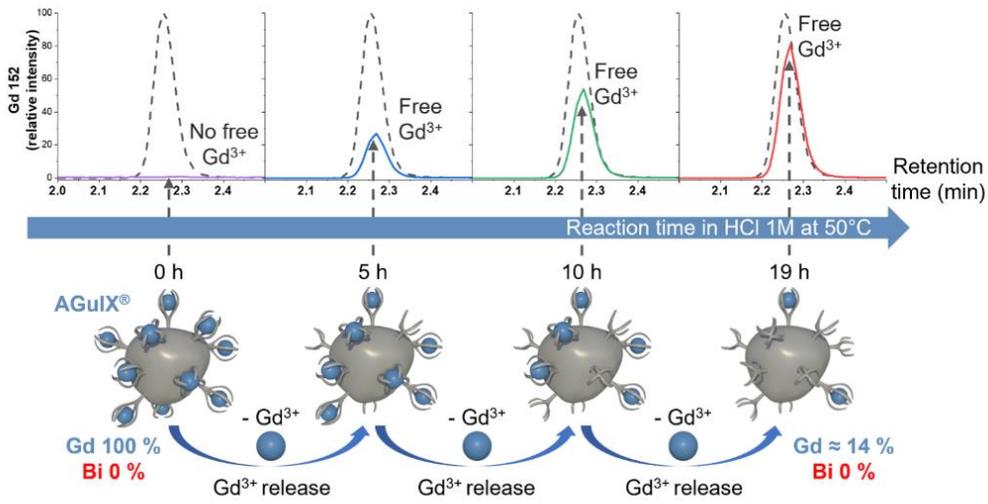
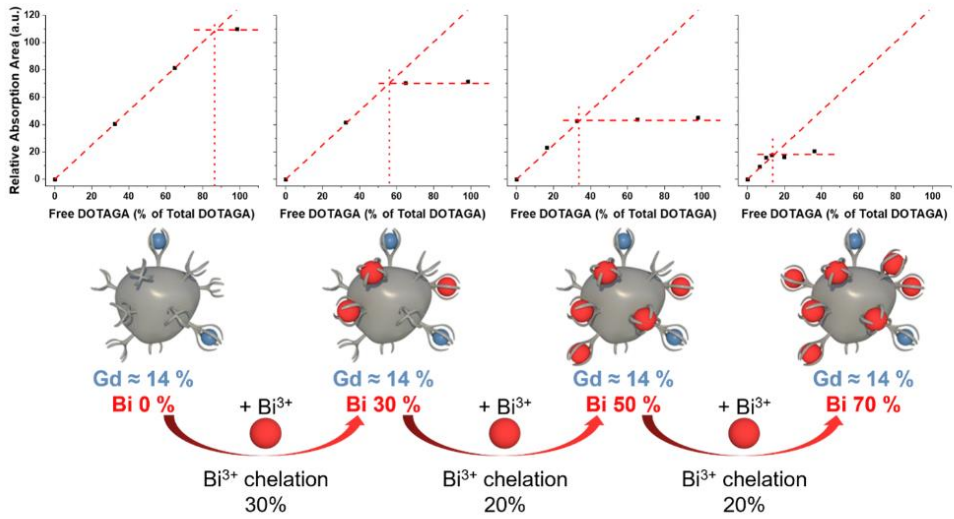
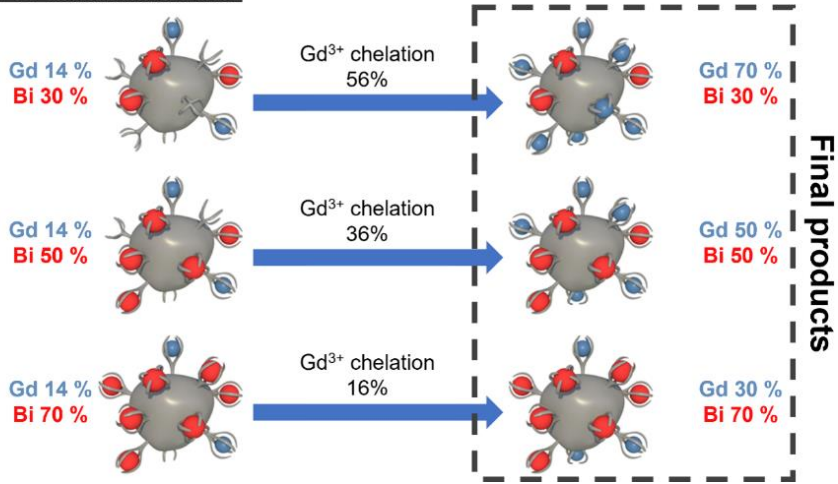

**Figure 1.** A) Increase of free gadolinium in the reaction mixture depending on the reaction time (in HCl 1 M at 50°C) followed by HPLC-ICP/MS and compared to gadolinium standard (dash plot) corresponding to 100% release of initial gadolinium. The lower part is a schematic view of the process starting from AGuIX® to obtain a particle devoid of about 80% of its initial gadolinium content. B) Results of the free DOTAGA titration at each step of the bismuth chelation show the diminishment of free DOTAGA amounts all along the chelation process. The lower part is an illustration of the bismuth chelation process beginning with 20Gd/0Bi, to generate 3 different batches of particles: 20Gd/30Bi, 20Gd/50Bi, and 20Gd/70Bi. C) Schematic view of the final gadolinium chelation to obtain the 3 final AGuIX-Bi products with desired Gd/Bi ratios: 70Gd/30Bi, 50Gd/50Bi, and 30Gd/70Bi (molar ratio).

**Characterization of AGuIX-Bi nanoformulations**

The size of a nanoparticle impacts its pharmacokinetics [32] and its elimination pathway [32, 33]. Dynamic light scattering (DLS) was used to measure the hydrodynamic diameter ($D_H$) of the NPs. Volume distributions of the $D_H$ for the three final products are displayed in **Figures S5-S7**. The measured $D_H$ for the three Gd/Bi ratios are similar: $4.5 \pm 1.6$ nm for 70Gd/30Bi, $5.1 \pm 1.9$ nm for 50Gd/50Bi, and $4.8 \pm 1.9$ nm for 30Gd/70Bi (**Table 1**). All NP formulations have diameters below the glomerular cutoff of < 8 nm [33], suggesting potential for renal clearance from the body similar to previously published AGuIX® NPs [34]. The overall hydrodynamic diameter of the AGuiX-Bi NPs is close to the diameter of AGuIX® NPs with a $D_H = 4.6 \pm 1.6$ nm (**Table 1**). The volume distribution is displayed in **Figure S8**.

The chromatograms (HPLC-UV, High-performance liquid chromatography-Ultraviolet) at 295 nm agree with the maximum absorption of the original AGuIX® [35] nanoparticle, as well as a high absorption area of DOTAGA(Bi) [36]. The retention times (Tr) of the three final products

70Gd/30Bi, 50Gd/50Bi, and 30Gd/70Bi respectively 10.6 min, 10.6 min, and 10.4 min; comparable to the AGuIX® retention time of 10.9 min (**Table 1**). These results confirm the low impact on final NP formulation characteristics, making clinical scalability of the NPs with various ratios of Gd and Bi more feasible as well as providing confidence in the ease of nanoformulation tunability to incorporate other heavy metals, drugs, or isotopes for additional multimodal diagnosis and therapeutics compared to the previous methodology [31]. A high purity of 92% is obtained for the three final products measured directly from the HPLC-UV chromatogram absorbance ratio at 295 nm ($\% \, purity = \frac{ABS_{295}\,[10-15\,min]}{ABS_{295}\,[0-15\,min]}$) (**Figure S9**). We observe an overall increase in peak absorbance for all nanoparticles compared to AGuIX®. This increase is due to the obtention of DOTAGA(Bi) complexes [36] and is proof of the chelation process efficiency. This efficacy was further confirmed through the HPLC-SEC-ICP-MS (SEC, size-exclusion chromatography) chromatogram (**Figure S10A**) where we observe the co-localization of Bi, Gd, and Si (Silicone) of the nanoparticle core. The Gd/Bi ratio evolution can also be seen through the ICP-MS intensity variation on each chromatogram (**Figure S10B**).

The ζ-potentials show that the pH is equal to 7.11 at the iso-electric point (IEP) (**Table 1**) and is the same for all three final products. As the three products come from the same starting material, those equal IEPs are a sign of the same final complexation levels. Moreover, their IEP is close to the IEP of the initial AGuIX® at 7.15 (**Table 1**), indicating that the surface of the three final products and the original AGuIX® particle should be similar. In addition, the release and the chelation process can also be followed by ζ-potential measurement and IEP change. As shown in **Figure S11**, the pH at IEP is decreasing after gadolinium release (5.58 vs. 7.15 for AGuIX®) which is a sign of the presence of free DOTAGA at the surface. During the Bi chelation step, we observe an increase after each step moving from 5.58 at the beginning of the process to 6.95 after the final Bi chelation. This increase is due to the formation of DOTAGA(Bi) at the particle surface.

An elemental analysis of the three final products can be found in **Table 1**. The Gd and Bi massic content of each final product is in perfect accordance with their desired Gd/Bi molar ratio. The product AGuIX-Bi (70Gd/30Bi) contains 479 µmol of Gd and 203 µmol of Bi per mg of product which corresponds to 70.3% of Gd and 29.7% of Bi. The product AGuIX-Bi (50Gd/50Bi) contains 342 µmol of Gd and 327 µmol of Bi per mg of product which corresponds to 51.1% of Gd and 48.9% of Bi. The product AGuIX-Bi (30Gd/70Bi) contains 201 µmol of Gd and 477 µmol of Bi per mg of product which corresponds to 29.6% of Gd and 70.4% of Bi. The three products have a similar total metal content of 682 µmol mg$^{-1}$, 669 µmol mg$^{-1}$, and 678 µmol mg$^{-1}$, respectively. This total metal is lower than the initial AGuIX® metal content equal to 744 µmol mg$^{-1}$ which can be the consequence of some hydrolysis of the amide bond linking the DOTAGA groups to the nanoparticle core (due to acidic conditions) resulting in a decrease of the total DOTAGA content or to the elimination of some smaller nanoparticles during the purification steps.

AGuIX-Bi (30Gd/70Bi) longitudinal relaxivity ($r_1$) at 7 T was 7 s$^{-1}$ Gd mM$^{-1}$ compared to 5 s$^{-1}$ Gd mM$^{-1}$ for AGuIX®. A similar trend was seen at 1.5 T with $r_1$ increasing from 18.7 s$^{-1}$ Gd mM$^{-1}$ for AGuIX® to 23.7, 20.1, 30.2 s$^{-1}$ Gd mM$^{-1}$ for AGuIX-Bi 70Gd/30Bi, 50Gd/50Bi, and 30Gd/70Bi respectively. Similar to previous literature for Gd$^{3+}$ compounds, $r_1$ decreased with increasing magnetic field strength. At higher field strengths, the paramagnetic property of a contrast agent is enhanced leading to stronger $T_2$-related signal decay and lower $r_1$ relaxivity [37]. Consequently, for clinical MR-Linac systems operating at even lower magnetic fields of 0.35 – 1 T, there is potential for further improved contrast for Gd$^{3+}$ compounds such as AGuIX-Bi compared to the 7 T and 1.5 T presented. A recent study comparing Gd$^{3+}$ concentration per nanoparticle and $r_1$ relaxivity demonstrated that at low Gd$^{3+}$ concentrations, $r_1$ relaxivity increased until a maximum was reached. Further addition of Gd$^{3+}$ caused a quenching effect with decreasing $r_1$ relaxivity relative to concentration [38]. This quenching effect was attributed to the packing of Gd$^{3+}$ into a limited volume, potentially restricting access of $H_2O$ molecules to the $Gd^{3+}$ coordination sphere. Similar quenching effects have been observed in other $Gd^{3+}$-based nanoparticle systems [39-41] and may explain the increase in relaxivity as the ratio of gadolinium within AGuIX-Bi was decreased thus reducing packing. Studies have also suggested that increased $Gd^{3+}$ payload may lead to disproportionately increased $T_2$ effects which could negatively impact $r_1$ relaxivity. The increased $T_2$ contribution is primarily attributed to $Gd^{3+}$ geometric confinement which increases dipolar interactions between neighboring gadolinium ions [38-43]. We observed a decrease in the $r_1/r_2$ ratios as the concentration of $Gd^{3+}$ increased suggesting a disproportional increase in $r_2$ relaxivity contribution. To confirm retention of MRI signal and compare differences in Gd/Bi ratios in clinically relevant MRI systems, agar phantoms of varying nanoparticle concentrations were imaged in a 0.35T MRIdian MRI-Linac (**Figure S12A**) and 3T Siemens MR scanner (**Figure S12B**). As expected AGuIX® maintained the strongest MRI signal followed by AGuIX-Bi (70Gd/30Bi), and equivalent signal for AGuIX-Bi (50Gd/50Bi) and AGuIX-Bi (30Gd/70Bi). Overall minimal loss of MRI signal was seen with increasing ratios of Bi confirming strong contrast was maintained even at the highest Bi concentration in multiple clinical scanners.

| Product | Gd (w%) | Bi (w%) | $D_H$ (nm) | pH IEP | 1.5 T | | | 7 T |
| --- | --- | --- | --- | --- | --- | --- | --- | --- |
| | | | | | $r_1$ ($s^{-1}$ Gd m$M^{-1}$) | $r_2$ ($s^{-1}$ Gd m$M^{-1}$) | $r_1/r_2$ | $r_1$ ($s^{-1}$ Gd m$M^{-1}$) |
| AGuIX® | 11.7 | - | 4.6 ± 1.6 | 7.15 | 18.7 | 30.3 | 1.62 | 5.0 |
| AGuIX-Bi (70Gd/30Bi) | 7.54 | 4.24 | 4.5 ± 1.6 | 7.11 | 23.7 | 41.4 | 1.75 | - |
| AGuIX-Bi (50Gd/50Bi) | 5.38 | 6.84 | 5.1 ± 1.9 | 7.11 | 20.1 | 36.0 | 1.79 | - |
| AGuIX-Bi (30Gd/70Bi) | 3.16 | 9.97 | 4.8 ± 2.0 | 7.11 | 30.2 | 56.7 | 1.88 | 7.0 |

**Table 1.** Physical and chemical characteristics comparison of AGuIX® and AGuIX-Bi 70Gd/30Bi, 50Gd/50Bi and 30Gd/70Bi products.

**Biological characterization of AGuIX-Bi**

Studies to assess toxicity and treatment efficacy were conducted in human NSCLC cells (A549), derived from a 58 y male epithelial carcinoma patient with a KRAS mutant and EGFR wild-type [44]. An MTT assay was used to assess mitochondrial metabolism as an indicator of nanoparticle toxicity and cellular viability. MTT reagent permeates live cell membranes and is reduced by oxidoreductase, dehydrogenase enzymes, and NAD(P)H (nicotinamide adenine dinucleotide phosphate) during different stages of the mitochondrial electron transport chain [45]. When comparing the cellular metabolism of the various AGuIX-Bi formulations (**Figure 2A**), minimum toxicity was seen up to 3 mg mL$^{-1}$ following 24 h post-incubation, similar to AGuIX® (**Figure S13**). We calculated logIC$_{50}$ values of 1.6, 1.5, 1.4, and 1.5 mg mL$^{-1}$ for AGuIX®, AGuIX-Bi (70Gd/30Bi), AGuIX-Bi (50Gd/50Bi), and AGuIX-Bi (30Gd/70Bi), respectively. Breaks in double-stranded DNA (dsDNA) following radiation were assayed using γH2AX, a well-established marker for double-strand breaks (DSBs); histone H2AX is phosphorylated by ATM (ataxia-telangiectasia mutated) at or near sites of DSBs to recruit repair machinery [46-48]. Quantification of DSBs is an indicator of nanoparticle radiation dose amplification and was evaluated for the different ratios of Gd/Bi nanoparticles (**Figure 2B, Figure S14A**) and AGuIX® (**Figure S14B**). As the ratio of Bi within the formulation increased, the number of DSBs increased at 4 Gy and 6 Gy confirming the predicted impact of including a higher Z atom within the formulation. This correlation between damage and an increase in the Bi ratio was further confirmed with clonogenic assays comparing the three AGuIX-Bi formulations (**Figure S15A**). Given that increased Bi content amplified damage without impacting cellular viability, all following biological characterizations were conducted with AGuIX-Bi (30Gd/70Bi) NPs.

Clonogenic assays assessed the radiation dose amplification efficacy of AGuIX® versus AGuIX-Bi (30Gd/70Bi) confirming decreased viability with a sensitizer enhancement ratio (SER$_{50\%}$) of 1.28 for AGuIX-Bi (30Gd/70Bi) (**Figure 2C, Figure S15B**). Although SER values can be heavily impacted by various confounding factors including concentration and cellular uptake of sensitizer, tumor model, and radiation energy (keV vs MeV) and source (photons vs protons), an SER of 1.28 is comparable to other clinical and experimental radiosensitizers such as gemcitabine [49, 50], hafnium oxide nanoparticles [51], and gold nanoparticles [52, 53]. Further comparison of AGuIX® and AGuIX-Bi (30Gd/70Bi) DNA damage was measured directly using the alkaline CometChip assay (**Figure S16**) following 6 Gy and 10 Gy radiation. Damage was quantified as the percentage of DNA in the comet tail region. Higher levels of basal DNA damage were observed in mock-irradiated cells treated with NPs, consistent with previous reports [25] (**Figure 2D**). At 10 Gy, a higher level of DNA damage was induced with AGuIX-Bi (30Gd/70Bi) compared with AGuIX®. The alkaline CometChip assay thus corroborates the finding of enhanced radiation-induced damage with the addition of Bi to the nanoparticle formulation. To determine how NP and radiation treatment would impact cell death, we co-treated cells with either AGuIX® or AGuIX-Bi (30Gd/70Bi) and ionizing radiation at 6 Gy and 10Gy to measure apoptosis induction after 24 and 48 hours. We found that AGuIX-Bi (30Gd/70Bi) at 10 Gy induced more apoptosis compared to AGuIX® (**Figure 2E**). Also, we noticed cells became positive for annexin V (AxV), which binds to externalized phosphatidylserine in the early stages of apoptosis, before becoming positive for propidium iodide (PI), which only enters cells in the later stages of apoptosis when the plasma membrane is compromised [54] (**Figure S17**). This indicates that cells co-treated with these nanoparticles and ionizing radiation were undergoing apoptotic cell death. Interestingly when comparing DNA damage and clonogenic survival of AGuIX-Bi (30Gd/70Bi) to AGuIX® at the same massic concentration, AGuIX-Bi (30Gd/70Bi) had more DNA damage and apoptosis induction even though the total molar metal content is 9.7% lower than AGuIX®, further highlighting the impact of the higher atomic number of Bi on the photoelectric cross-section and subsequent therapeutic effect. No differences in nanoparticle cellular internalization were identified in NSCLC (**Figure 2F and Figure S18**) cells suggesting the decrease in cellular viability was due to the addition of high Z Bi rather than changes in nanoparticle uptake.

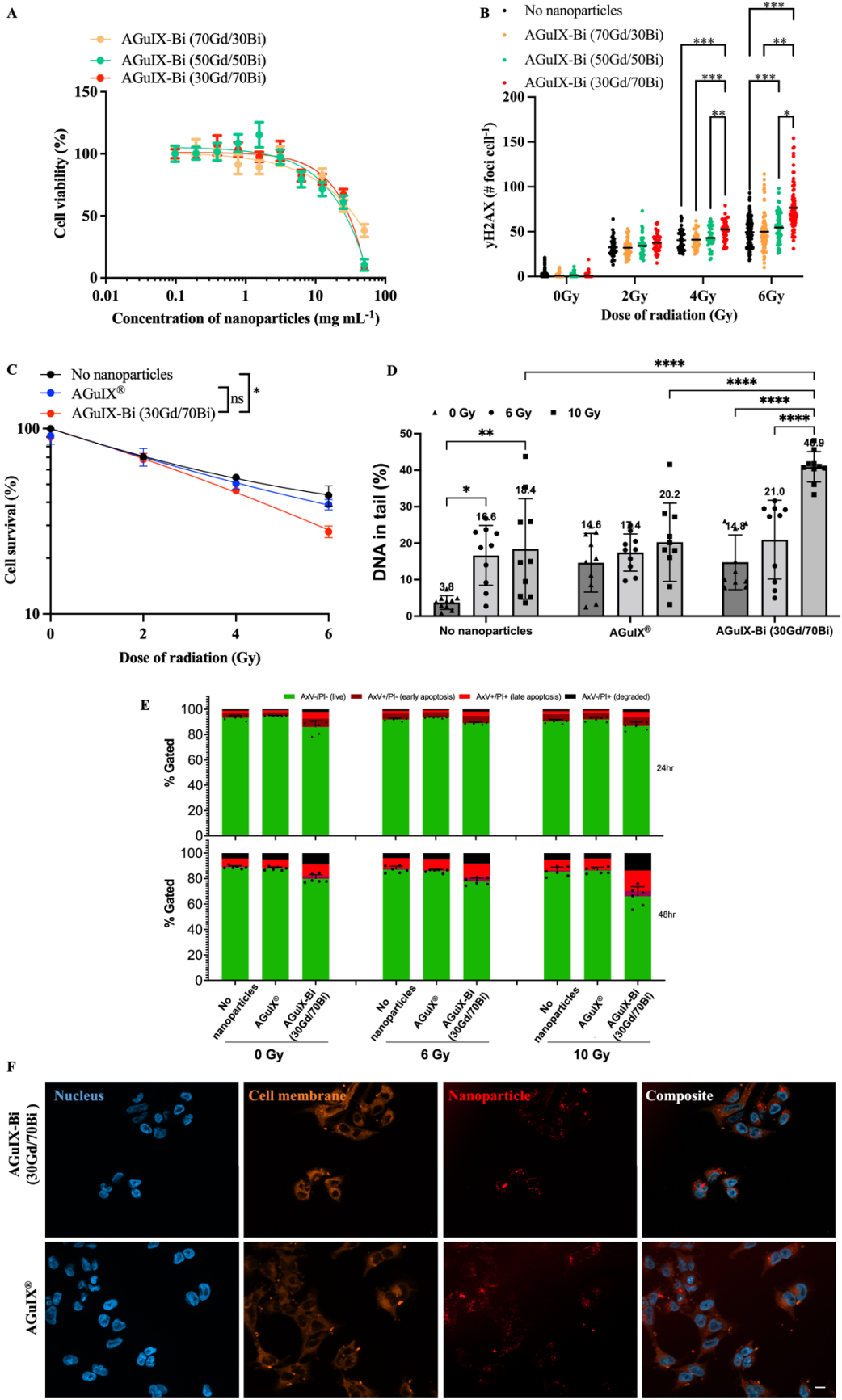

**Figure 2.** A) *In vitro* cytotoxicity of AGuIX-Bi nanoformulations as a function of concentration 24 h post-incubation of varying ratios of 70Gd/30Bi, 50Gd/50Bi, and 30Gd/70Bi in a NSCLC A549 murine model. For all three formulations, particles were tolerated up to 3 mg mL$^{-1}$ (n = 3). B) Breaks in dsDNA following radiation was assayed using γH2AX which indicated that radiation dose amplification damage correlated with increasing ratios of Bi (n = 60-80 cells). Data presented as mean ± SEM (* $p< 0.05$, ** $p<0.01$, and *** $p<0.001$). C) Clonogenic assay comparing the long-term survivability of A549 NSCLC in the presence of either AGuIX® or AGuIX-Bi (30Gd/70Bi) and radiation therapy (n = 3). Data presented as mean ± SEM. (* $p< 0.05$) D) DNA damage induced by AGuIX® or AGuIX-Bi (30Gd/70Bi) treated A549 cells irradiated at 6 Gy (circle symbol) or 10 Gy (square symbol) or sham-irradiated (0 Gy, triangle symbol). DNA in the tail region is plotted on the y-axis against different NPs treatments on the x-axis. Each symbol represents the median value of DNA in the tail region for at least 50 comet cells per well. Data presented as mean ± SD (* $p<0.05$, ** $p<0.001$, **** $p<0.0001$). E) Chemosensitivity assay to study induction of A549 cell death via apoptosis following treatment with either AGuIX® or AGuIX-Bi (30Gd/70Bi) and 6 Gy or 10 Gy irradiation (n = 6). Cells were stained for the presence of externalized phosphatidylserine with annexin V (AxV) and for plasma membrane disruption with propidium iodine (PI). Gating presented: Green = AxV-PI- (live cells), Dark purple = AxV+PI- (early apoptosis), Red = AxV+PI+ (late apoptosis), and Black = AxV-PI+ (degraded). F) Qualitative internalization representative microscopy images at equivalent massic concentrations (1 mg mL$^{-1}$) of either AGuIX® or AGuIX-Bi (30Gd/70Bi). Blue = DAPI stained nucleus, Orange = cell membrane, and Red = Cy5.5 tagged NPs, Scale bar = 5 µm.

Biodistribution studies in NSCLC implanted mice compared AGuIX® accumulation and clearance to AGuIX-Bi (30Gd/70Bi) via ICP-MS and MRI. Although AGuIX-Bi (30Gd/70Bi) (**Figure 3A**) and AGuIX® (**Figure 3B**) had similar physio-chemical characteristics (size and surface charge), biodistribution and pharmacokinetics varied surprisingly. ICP-MS identified the presence of Gd and Bi in bulk organs. Controls of non-injected mice were assessed to confirm positive elemental analysis was due to nanoparticle accumulation rather than tissue background. AGuIX-Bi (30Gd/70Bi) relative injected dose within all organs including tumors was higher compared to AGuIX®, likely due to increased blood circulation time. A Bi/Gd molar ratio of $2.3 \pm 0.56$ was observed across all extracted tissues, aligning with the predicted theoretical value of 2.33, thus suggesting *in vivo* Gd/Bi co-localization of AGuIX-Bi (30Gd/70Bi) NPs (**Figure S19**). Further confirmation of *in vivo* stability needs to be considered in future translational work. A recent study [55] assessed the degradation of AGuIX® in PBS, human blood serum, and urine to determine if free gadolinium is released, which is considered harmful for patients [56]. This study confirmed that the primary product of AGuIX® degradation is not free Gd but rather APTES-DOTAGA-Gd (APTES, (3-Aminopropyl)triethoxysilane), a chelated form of Gd. Given that AGuIX-Bi (30Gd/70Bi) is a derivative of the AGuIX® nanoparticles which have demonstrated success in phase I dose escalation studies and ongoing phase II clinical trials, we expect similar chelated degradation products that minimize risk to patients.

AGuIX® had high tumor accumulation 1h post-injection followed by rapid clearance, while AGuIX-Bi (30Gd/70Bi) was retained within the tumor 24 h post-injection (**Figure 3C**). AGuIX® had a shorter blood circulation time with rapid clearance between 1-4 h while AGuIX-Bi (30Gd/70Bi) had slower kinetics and high blood concentrations persisted 4 h post-injection with clearance finally observed at 24 h. Both AGuIX® and AGuIX-Bi (30Gd/70Bi) were primarily eliminated via renal clearance (40 - 60% ID $g^{-1}$), however 10 - 15% ID $g^{-1}$ of AGuIX-Bi (30Gd/70Bi) cleared through the liver compared to only 2 - 3% ID $g^{-1}$ of AGuIX®. Nanoparticles sub-5 nm can quickly clear via renal excretion as seen with AGuIX®, however, a longer circulation time is associated with increased surface serum protein accumulation which enhances opsonization and phagocytosis via the mononuclear phagocyte system (MPS) associated with increased liver and splenic filtration [57, 58]. AGuIX® and AGuIX-Bi (30Gd/70Bi) MRI tumor accumulation and retention mirrored ICP-MS biodistribution results with AGuIX® peaking 30 mins post-injection and AGuIX-Bi (30Gd/70Bi) peaking at 30 mins and sustaining accumulation for 24 h (**Figure 3D**). AGuIX® accumulation correlated with previously reported work in orthotopic lung, pancreatic, hepatic, and brain cancer models supporting EPR-based preferential accumulation [17, 59, 60]. Although there is a difference in peak contrast enhancement for AGuIX® compared to AGuIX-Bi (30Gd/70Bi) at 30 mins post-injection, this variance can be attributed to the difference in total Gd content in each NP (**Figure 3E, Figure S20**), specifically, AGuIX® has approximately 3-fold more Gd per nanoparticle compared to AGuIX-Bi (30Gd/70Bi). Even at the lower Gd content, sufficient T1-weighted contrast was present to identify the tumor compared to surrounding healthy tissue for AGuIX-Bi (30Gd/70Bi). To further explore the differences in biodistribution, Taylor Dispersion Analysis (TDA) ICP-MS was used to better characterize the mean hydrodynamic radii of AGuIX-Bi (30Gd/70Bi) nanoparticles (**Figure S21**). Although DLS is a common technique to determine the hydrodynamic diameter of nanoparticles, the scattered light is proportional to the sixth power of the radius, thus often strongly biasing the results towards a larger particulate presence [61]. Compared to DLS, TDA can better resolve smaller species and is not biased by aggregates or any larger species that may be present. Previous work using TDA-ICP-MS confirmed that the average size of AGuIX® nanoparticles is $5.6 \pm 0.2$ nm ($87 \pm 5\%$) [55]. The same protocol was applied to AGuIX-Bi (30Gd/70Bi) confirming an increase in the overall size with a diameter of $7.1 \pm 0.3$ nm ($86 \pm 1\%$). This increase in size may have contributed to the longer blood circulation, tumor retention, and difference in clearance of the AGuIX-Bi (30Gd/70Bi) compared to AGuIX®. We predict this increase in size may be due to the preferential elimination of smaller nanoparticles during the acidic Gd release and subsequent purification steps, however further studies will be necessary to confirm this hypothesis. It should be noted that in both sets of data, we see a second population of smaller diameter particles (1.2 ± 0.7 nm, 13% AGuIX® and 1.7 ± 0.1 nm, 14%, AGuIX-Bi (30Gd/70Bi)), which may correspond to nanoparticle fragments due to the biodegradability of the nanoparticle in a highly diluted medium [34]. Based on ongoing clinical trials with AGuIX®, this smaller population does not seem to impact overall efficacy or toxicity and is eliminated first and rapidly as confirmed by extensive animal studies on AGuIX® NPs [34].

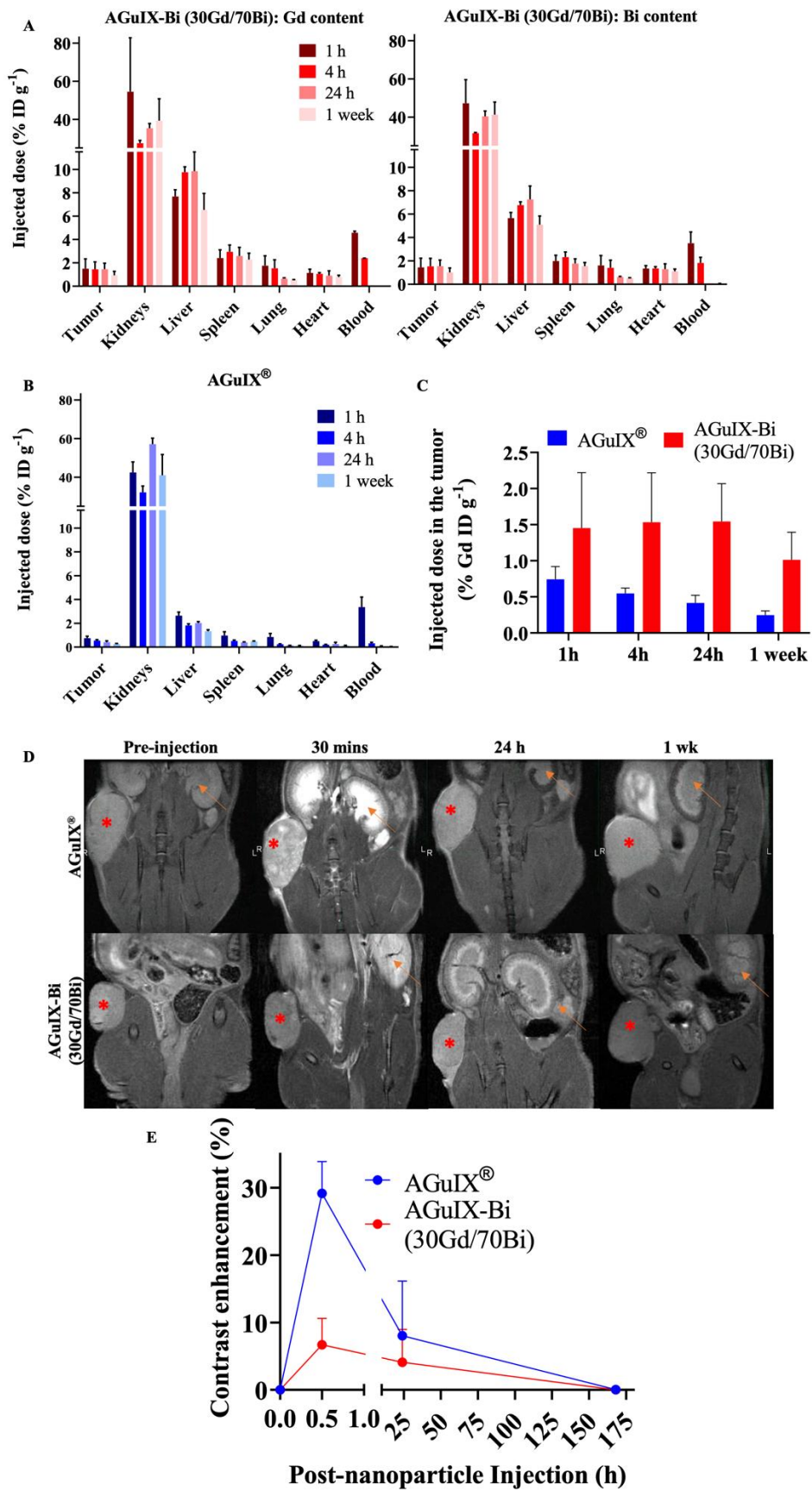

**Figure 3.** Biodistribution performed by ICP-MS in three animals per time point after retroorbital, intravenous injection of 300 mg kg$^{-1}$ of either A) AGuIX-Bi (30Gd/70Bi) or B) AGuIX® in a NSCLC A549 murine model. C) Tumor-specific retention of nanoparticles varied with AGuIX® clearing as early as 1h post-injection and AGuIX-Bi (30Gd/70Bi) maintaining sustained tumor nanoparticles for 24 h. Data is expressed as percent of the injected dose per gram of tissue (%ID g$^{-1}$) D) Qualitative representations of A549 subcutaneous mice tumor accumulation as a function of time. Reconstructed MR images were bias field corrected. A red asterisk indicates a tumor and an orange arrow indicates a kidney. E) Quantitative measure of contrast enhancement following nanoparticle injection relative to surrounding healthy muscle tissue (n = 3). Data shown as mean ± SEM.

## AGuIX-Bi toxicity and therapeutic efficacy

Short-term and long-term toxicity of AGuIX-Bi (30Gd/70Bi) was assessed via histopathology and serum toxicology. Short-term (**Figure 4A & S22-S23**) and long-term (**Figure S24**) toxicity were evaluated 24 h and 1 mo post-treatment, respectively. No evidence of toxicity was identified in any treatments, which was further confirmed by the evaluation of histopathology slides by a trained pathologist. Differences in histology, cell size, and color intensities among the different treatments were related to differences in tissue processing and fixation. Serum toxicity was evaluated 24 h post-radiation therapy for blood urea nitrogen (BUN), an indicator of kidney dysfunction since primary clearance of AGuIX-Bi (30Gd/70Bi) occurs via the kidney (**Figure 4B**). Evaluation of Alanine aminotransferase (ALT), Aspartate transaminase (AST), and Gamma-Glutamyl transferase (GGT) indicated no liver damage in treated mice. Compared to AGuIX®, AGuIX-Bi (30Gd/70Bi) had a significantly higher clearance of NPs via the liver, but no additional damage was identified. The systemic inflammatory response was assayed via Lactate dehydrogenase (LDH) to assess the impact of injecting foreign nanoformulations, however, no indication of damage was determined. Due to the difference in metabolic rates of mice versus humans, a direct comparison of dose and pharmacokinetics cannot be accurately measured [62], however the dose of 300 mg kg$^{-1}$ used in this study was three times higher than the clinically safe dose of 100 mg kg$^{-1}$ used in the NANO-RAD (NCT02820454) phase I dose escalation clinical trial for AGuIX®[63] and comparable to AGuIX® dose used in other animal models [14, 17, 64] showing no signs of toxicity.

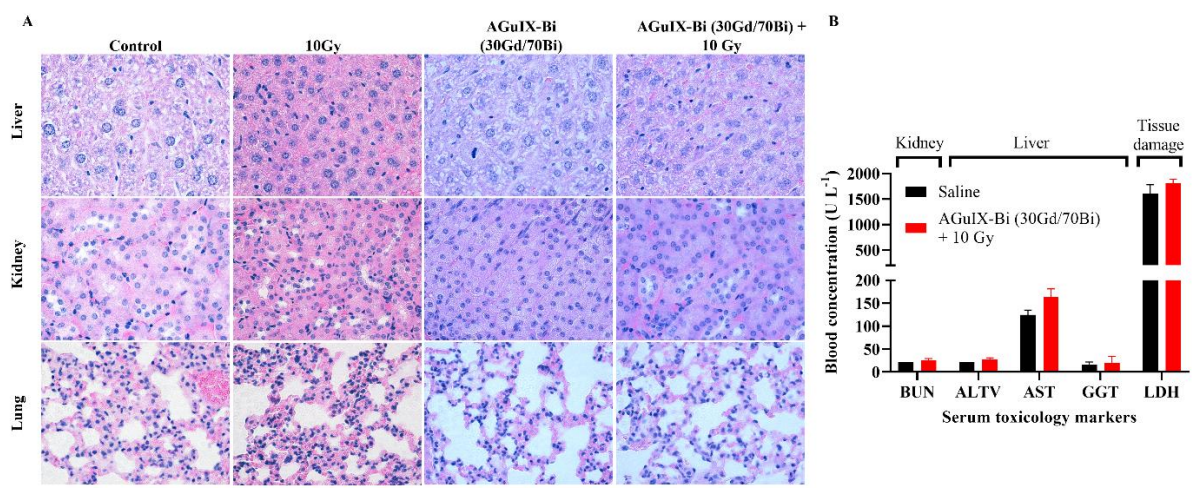

**Figure 4.** A) Histopathology of mice treated with either saline or AGuIX-Bi (30Gd/70Bi) and/or 10 Gy radiation 24 h post-treatment. No short-term toxicity was identified in the liver, kidney, or lungs. B) Blood chemistry analysis of serum 24 h post-treatment showed no signs of systemic kidney or liver dysfunction. (n = 3). Data shown as mean ± SEM. BUN = Blood urea nitrogen, ALT = Alanine aminotransferase, AST = Aspartate transaminase, GGT = Gamma-Glutamyl transferase, and LDH = Lactate dehydrogenase.

Therapeutic studies with radiation dose effect amplification were performed at the optimum time when NP tumor accumulation was high relative to healthy tissue, taking into account the substantial difference in biodistribution and pharmacokinetics for each NP. Optimum AGuIX-Bi (30Gd/70Bi) nanoparticle uptake occurred starting at 30 mins and was maintained for 24 h, however, nanoparticle blood concentration also remained high for the first 1-4 h with clearance at 24 h. To maximize tumor nanoparticle-mediated dose amplification and minimize non-specific toxicity, radiation was delivered 24 h post-AGuIX-Bi (30Gd/70Bi) injection. In contrast, optimum AGuIX® uptake occurred between 30 mins – 1 h with rapid tumor clearance as early as 1 h post-injection, similar to previously established literature in various murine tumor models [14, 17, 64, 65]. In the NANO-RAD (NCT02820454) phase I clinical trial, AGuIX® plasma half-life was determined to be 1.3 h (range 0.8-3 h) [63]. To compare the efficacy of AGuIX® at peak tumor accumulation and mirror previous literature treatment plans [14, 17, 64, 65], radiation was delivered 30 mins post-injection. Mice treated with AGuIX-Bi (30Gd/70Bi) and radiation therapy had significantly delayed tumor growth (**Figure 5A**) with 33% complete regression (**Figure 5B**). Mice treated with AGuIX-Bi (30Gd/70Bi) plus radiation had delayed mean tumor growth of 58 days compared to 22 and 18 days for AGuIX® plus radiation and radiation-only groups, respectively (**Figure 5C**). When combined with radiation therapy, both NPs maintained tumor growth delay up to day 27 post-treatment compared to the radiation-only group but eventually succumbed to exponential regrowth, indicating the potential need for repeat cycles of treatment to better mimic clinical treatment workflows that include administration of radiation therapy. Given that a single treatment dose had no impact on mouse weight indicating no systemic toxicity with either NP alone or in combination with radiation therapy (**Figure 5D**), multiple cycles of NP and radiation can be an option for sustained tumor delay and eventual regression. In the NanoCOL Phase Ib (NCT03308604) clinical trial, AGuIX® NPs were delivered two times during the 5 weeks of conventional radiation therapy (45 Gy, 25 sessions) and once again before two weeks of brachytherapy (15 Gy) for effective clinical impact in cervical cancer patients [14].

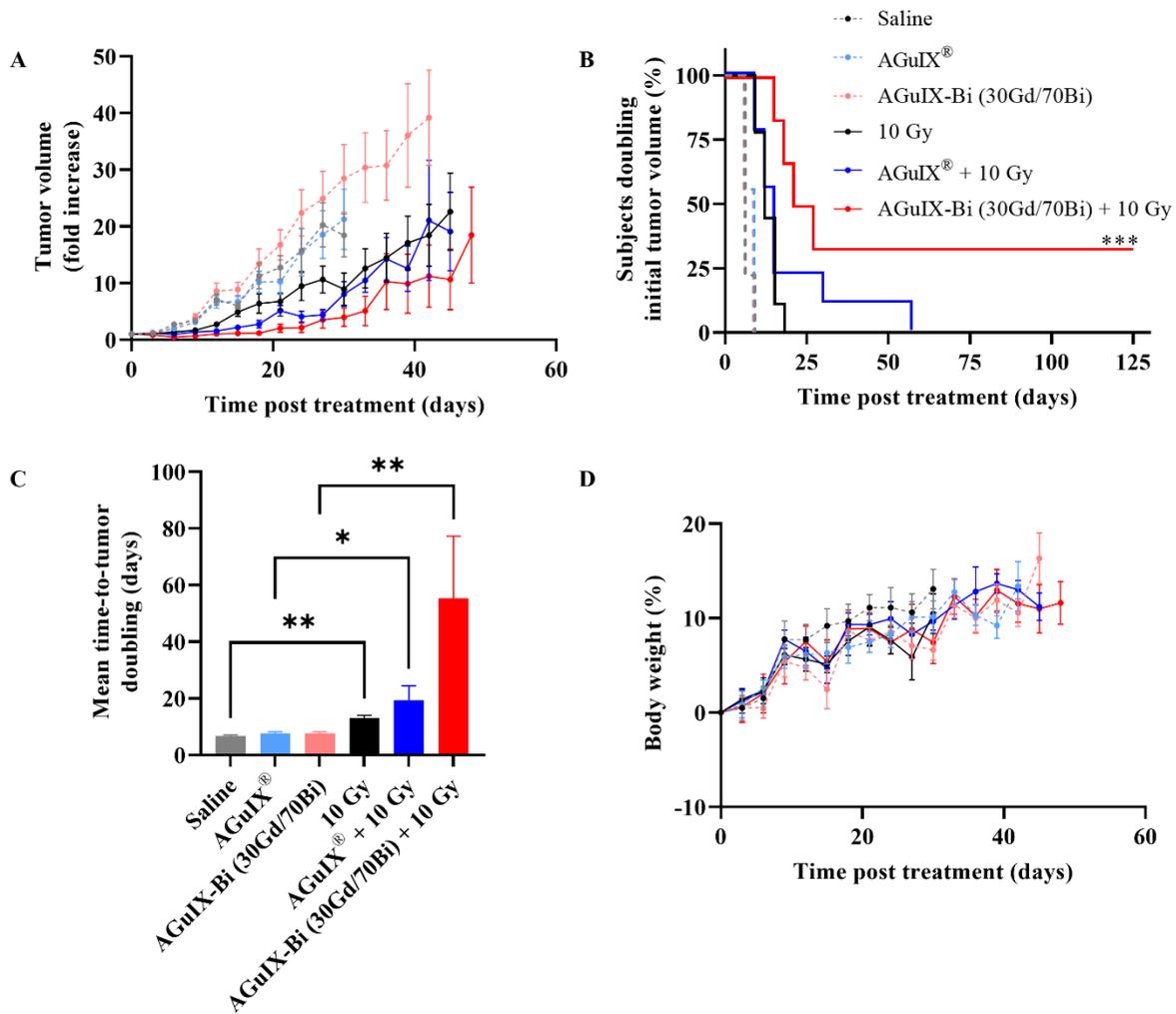

**Figure 5.** A) *In vivo* therapeutic efficacy of AGuIX® or AGuIX-Bi (30Gd/70Bi) with or without radiation (10 Gy) in a NSCLC A549 murine model. Tumor growth relative to the tumor size on the day of irradiation is shown (n = 6 – 9 mice). Nanoparticles were given at a massic concentration of 300 mg kg$^{-1}$ followed by radiation 30 mins post-AGuIX® or 24 h post-AGuIX-Bi (30Gd/70Bi) injection. Relative tumor growth for each group is presented until the day the first animal within that group dropped out. B) Time-to-tumor-doubling is indicated with endpoint failure defined as tumor volume doubling relative to Day 0. Complete tumor regression was observed in 33% of mice treated with AGuIX-Bi (30Gd/70Bi) plus radiation. C) Mean time-to-tumor doubling shows that mice treated with AGuIX-Bi (30Gd/70Bi) plus radiation had significantly longer tumor delay compared to control groups. D) No change in mouse body weight was observed following any treatment indicating no systemic toxicity. (* $p< 0.05$, ** $p<0.01$, and *** $p<0.001$)

In clinical trials, radiation is delivered 4 h post-AGuIX® injection allowing sufficient time for peak tumor accumulation (plasma half-life 1.3 h, range 0.8-3 h) while minimizing the impact on healthy tissues [14, 63]. However, counterintuitively a recent study in a glioblastoma rat model compared radiation 1 h and 24 h post-AGuIX® injection and found better efficacy at 24 h even though the total AGuIX® content within the tumor was 36x lower compared to 1 h. It was suggested that rapid clearance of the bulk of the nanoparticles between 1 - 24 h indicated weak interactions between the cancerous cells and the NPs, however AGuIX® that remained within the tumor at 24 h were allowed sufficient time to interact with the cells and adhere to the surface or even be internalized [64]. As previously stated, secondary photoelectric and Auger electrons have a short tissue interaction distance [21] suggesting that smaller distances from cells and specifically cellular nuclei may sustain a stronger radiosensitization impact at even lower doses. The longer tumor retention of high concentrations of AGuIX-Bi (30Gd/70Bi) suggests the possibility of greater clinical impact due to the potential for stronger interactions of nanoparticles and the surrounding cancerous tissue.

**Conclusion**

The idea of heavy metal nanoparticles as radiation dose amplifiers has been explored preclinically for over four decades with little advancement due to challenges in scalable cGMP formulation processes, desirable pharmacokinetics, and concerns of non-specific toxicity. In this work, we have developed a formulation methodology to synthesize next-generation theranostic bismuth-gadolinium nanoparticles using cGMP AGuIX®, which are currently in clinical trials, as a tunable chelation platform for ease of clinically scalable production. In a NSCLC model, we have demonstrated the impact of adding higher Z $Bi^{3+}$ to the nanoformulation leading to increased levels of DNA damage and decreased cellular viability. *In vivo,* we observed no short-term or long-term toxicity associated with the nanoformulation and improved therapeutic efficacy with 33% complete regression while maintaining MR contrast. Given the promise of AGuIX® in phase I and ongoing phase II clinical trials, these preliminary studies synthesizing and validating the theranostic application of AGuIX-Bi (30Gd/70Bi) are necessary to lay the foundation for the translation of these next-generation nanoparticles.

**Experimental Section/Methods**

**Chemical materials**

The precursor NPs AGuIX® were provided as a lyophilized powder by NH TherAguix (Grenoble, France). A 2 M HCl solution was created by diluting HCl (extra pure, 37%, CarlRoth). A 1 M NaOH solution was created from NaOH pellets (Fischer Chemical). The $Cu^{2+}$ titration solution was created by dissolving $CuCl_2$ (powder, 99%, Sigma Aldrich, France) in pH 2 aqueous solution. $BiCl_3$ and $GdCl_3(6H_2O)$ used for chelation were purchased at reagent grade (>98%, Sigma Aldrich). $HNO_3$ (69%, supra grade) used for organ digestion was purchased from CarlRoth. Acetonitrile ($CH_3CN$, ACN, > 99.9%) from Sigma Aldrich (France) and trifluoroacetic acid (TFA) from Fischer Chemical (USA) were used for the HPLC-UV phase preparation. TRIS, NaCl, and HCl obtained from Sigma were used for the preparation of the mobilization solution in TDA-ICP-MS.

**AGuIX® and AGuIX-Bi synthesis**

The AGuIX-Bi NPs have been synthesized starting from the original AGuIX® NPs. The synthesis has been performed in 3 steps. The first step consists of $Gd^{3+}$ release from the initial AGuIX® NPs to obtain free DOTAGA (2-(4,7,10-tris(carboxymethyl)-1,4,7,10-tetraazacyclododecan-1-yl) pentanedioic acid) ready to chelate new metal ions. The second step focuses on the $Bi^{3+}$ complexation of newly regenerated DOTAGA groups. The last step is the final $Gd^{3+}$ complexation to obtain the desired products with three different Gd/Bi molar ratios respectively 70Gd/30Bi, 50Gd/50Bi, and 30Gd/70Bi.

*Step 1 – $Gd^{3+}$ release*

AGuIX® NPs were dispersed in 200 g $L^{-1}$ in Milli-Q $H_2O$ ($\rho > 18$ M$\Omega$ $cm^{-1}$). The AGuIX® solution was combined with a HCl (2 M) solution 30 mins after dispersion leading to a final composition of 1 M HCl and 100 g $L^{-1}$ AGuIX®, which was left stirring at 50°C. The mixture was sampled to assess the $Gd^{3+}$ release by HPLC-UV and ICP-MS. The amount of released $Gd^{3+}$ was compared to a standard $Gd^{3+}$ reference corresponding to the initial $Gd^{3+}$ content. As different Gd/Bi ratio NPs will be synthesized (70Gd/30Bi, 50Gd/50Bi, 30Gd/70Bi), $Gd^{3+}$ release is pushed to a sufficient level to produce a product ready to create the 3 different final ratios. In our case, the release was pushed to at least 70% to obtain the product 30Gd/70Bi. After 19 h of reaction, the released level was reached. The mix was isolated and the free $Gd^{3+}$ was removed via purification with a Vivaflow 200 ultrafiltration system (PES, molecular weight threshold = 5 kDa, Sartorius Stedim Biotech, Germany). The final solution pH was adjusted to 7.4 by NaOH (1 M) and stored overnight at 50°C to ensure the complexation of any possible free $Gd^{3+}$ remaining in the medium. Then a measure of the free DOTAGA amount through $Cu^{2+}$ titration was performed to confirm the presence of sufficient amounts of free DOTAGA and preparation of the following chelation steps.

*Step 2 – $Bi^{3+}$ chelation*

The bismuth complexation is challenged by the low solubility product of Bi ($K_s$ (BiOH$_3$) = 3.2 x 10$^{-40}$)[66] and fast precipitation into BiOH$_3$, resulting in a white solution (**Figure S25**). To ensure a good complexation, this part was divided into 3 steps. In the first step, 30% of the total DOTAGA amount was chelated. It was followed by two 20% complexation steps to reach 70% of DOTAGA chelated with Bi$^{3+}$. After each Bi$^{3+}$ complexation step, a part of the mixture was isolated to carry out Gd$^{3+}$ complexation and obtain the corresponding final product with the expected Gd/Bi ratio. This second step ended with 3 different products having different Gd/Bi ratios equal to 20Gd/30Bi, 20Gd/50Bi, and 20Gd/70Bi.

During a complexation step, the required amount of BiCl$_3$ was added to the mixture, then the pH was adjusted to 6-7 and the mixture was placed at 80°C under stirring. The mixture was left at 80°C for two days with regular pH readjustment to 6-7. After two days, the initial precipitated white solution was clear (**Figure S25**) and the pH was stable. To ensure the end of the complexation step, a sample of the mix was extracted to perform a DOTAGA titration.

*Step 3 – Gd$^{3+}$ complexation*

Each batch with their respective amount of Bi (30%, 50%, 70%) followed a Gd$^{3+}$ complexation step to reach the final expected Gd/Bi proportion in the particle: 70Gd/30Bi, 50Gd/50Bi and 30Gd/70Bi. The process was the same as for step 2, except that the right amount of GdCl$_3$(6H$_2$O) was added to the mixture taking into account the already chelated Gd. In the end, each solution was purified with the Vivaflow 200 ultrafiltration system to remove the excess chloride and sodium ions. Finally, the solution was sterile filtered through a 0.2 µm filter, removing any large impurities. NPs were freeze-dried using the Christ-Alpha 1-2 lyophilizer and stored.

**HPLC-ICP/MS to follow gadolinium release**

Throughout the process, reaction mixture samples were analyzed to follow the Gd$^{3+}$ release. Analysis was performed with a Perkin - Elmer Nexion 2000B combined with a Perkin-Elmer

Flexar LC. Separation was accomplished by a C4 reverse phase HPLC column (Jupiter®, 5 µm, 300 A, 150 x 4.6 mm) and measurements were conducted in isocratic mode using the phase composition H$_2$O / ACN / TFA (98.9 / 1 / 0.1% v) at 1 mL min$^{-1}$ speed flow. Gd signal was monitored using $^{152}$Gd. The operating conditions for the ICP-MS were as follows: auxiliary gas flow rate = 1.2 L min$^{-1}$; plasma gas flow rate =15 L min$^{-1}$; nebulizer gas flow rate = 0.84 L min$^{-1}$; RF power = 1600 W for the plasma. The remaining parameters were tuned for maximization of the Gd signal. Syngistix software (version 2.3) controlled the ICP-MS and Empower (version 7.3) was used to acquire the Gd signal.

**Measurement of gadolinium and bismuth in the final product by ICP-AES**

The measure of Gd and Bi amount in the final products was conducted by Crealins laboratory from the 6NAPSE group (ISA, Lyon, France). The measurements were performed through ICP-AES iCAP 65000 Duo device (Thermo Scientific).

**HPLC-UV**

Gradient HPLC analysis was performed using the Shimadzu Prominence series UFLC system with a LC-20 AD liquid chromatograph, a CBM-20A controller bus module, a CTO-20A column oven, and an SPD-20A UV-visible detector. UV-visible absorption was measured at 295 nm. Separation was performed at a flow rate of 1 mL min$^{-1}$ using C4 reverse phase HPLC column (Jupiter®, 5 µm, 300 A, 150 x 4.6 mm). The gradient initial solution was 95% solvent A – 5% solvent B (A = H$_2$O / ACN / TFA: 98.9 / 1 / 0.1% v, B = ACN / H$_2$O / TFA: 89.9 / 10 / 0.1% v) over 5 min. During the next step, samples were eluted over 15 mins by a gradient developed from 5 to 90% of solvent B in solvent A. The concentration of solvent B was maintained over 5 min and then the concentration of solvent B was decreased to 5% over a period of 5 min to re-equilibrate the system, followed by an additional 5 min at this final concentration.

**Free DOTAGA measurement**

Measurement of the amount of free DOTAGA groups was performed at different steps of the process either to evaluate the initial amount or to follow the complexation steps. Samples with increasing amounts of copper ($Cu^{2+}$) were prepared in acetate buffer (pH 4.5) and left to react for 30 min before being analyzed by HPLC-UV (method above). Free DOTAGA groups were measured by recording the increase in absorbance at 295 nm due to the formation of DOTAGA@($Cu^{2+}$) during titration. The amount of uncomplexed DOTAGA was directly related to the breaks in the slope of the absorbance graph.

**ζ-potential and dynamic light scattering (DLS) measurements**

NPs were diluted in an aqueous solution containing 0.01 M NaCl to 10 g $L^{-1}$ and size distribution along with ζ-potential was measured with a Zetasizer NanoS DLS (Dynamic Light Scattering, laser He-Ne 633 nm, Malvern Instrument).

**Relaxivity Measurements**

Relaxivity measurements were performed on a Bruker Minispec mq60 NMR analyzer (Bruker, USA) at 37°C, 1.5 T (60 MHz) with NPs at a concentration of 100 or 50 g $L^{-1}$.

**Agar phantoms clinical MRI**

Agar phantoms (1%) of various concentrations of AGuIX® or AGuIX-Bi (70Gd/30Bi, 50Gd/50Bi, or 30Gd/70Bi) were prepared by dissolving noble agar (Sigma-Aldrich) in boiling deionized water. While still liquid, NPs were added to 1 mL of agar solution in microvials and allowed to cool and solidify creating phantoms with concentrations of 0.0039 – 2 mg $mL^{-1}$ NPs. Phantoms were imaged using a $T_1$-weighted clinical sequence on the 0.35T MRIdian MRI-Linac and 3T Siemens MR. Images were quantified using 3D Slicer.

**Cell culture studies**

Human non-small cell lung cancer cells (A549, ATCC CCL-185, Manassas, VA) were cultured in 10% fetal bovine serum (FBS; Invitrogen, USA) and 1% pen/strep (10,000 U $mL^{-1}$ penicillin;

10,000 µg mL$^{-1}$ streptomycin; Invitrogen, US) supplemented Roswell Park Memorial Institute medium (RPMI 1640; Gibco, Invitrogen, USA) at 37°C, 5% CO$_2$ and optimal humidity.

**MTT cell viability assay**

A549 cells (10,000 cells well$^{-1}$) were seeded in a 96-well plate and allowed to grow for 24 h. Varying concentrations of each formulation (0 – 25 mg mL$^{-1}$) of AGuIX-Bi or AGuIX® NPs were diluted in 100 µL culture media and incubated with cells for 24 h. Cells were then washed and 100 µL of 10% MTT solution (5 mg mL$^{-1}$) was added and incubated for 4 h at 37°C. MTT crystals were dissolved in 100 µL of DMSO and absorbance was quantified at 570 nm.

**Clonogenic assay**

A549 cells were seeded at 250 – 1000 cells per well and incubated with 1 mg mL$^{-1}$ AGuIX® or AGuIX-Bi (70Gd/30Bi, 50Gd/50Bi, or 30Gd/70Bi) for 24 h. Following incubation, cells were irradiated at 2, 4, and 6 Gy using a 220 kVp Small Animal Research Radiation Platform (SARRP, Xstrahl, Suwanee, GA) at a dose rate of 4 Gy per minute. The cells were allowed to grow for approximately 7 – 10 days or until control well colonies reached 50 cells per colony. Cells were fixed with 10% neutral buffered formalin overnight and stained with 1% crystal violet overnight. Quantification was conducted using an in-house ImageJ (v 1.52p) colony counter macro. The sensitizer enhancement ratio (SER) was calculated as the ratio of doses at 50% survival fraction with and without nanoparticles ($SER_{50\%} = \frac{D_{NP+Rad}}{D_{Rad}}$).

**γH2AX assay**

Cells were seeded at a density of 10,000 cells on microscopy slides overnight. A549 cells were incubated with each nanoparticle formulation (1 mg mL$^{-1}$) overnight followed by irradiation on the SARRP to 2, 4, and 6 Gy. One hour post-treatment, cells were fixed, blocked, and permeabilized with 10% FBS and 0.3% Triton X-100. Cells were stained with anti-γH2AX-AlexaFlour 488 overnight and then mounted with DAPI and Flouromount-G. Images were taken using a Zeiss

AxioObserver microscope 100x and analyzed using an in-house ImageJ (v 1.52p) foci counter macro (**Figure S26**).

**Alkaline CometChip assay**

To quantitate the amount of DNA damage induced by AGuIX® or AGuIX-Bi (30Gd/70Bi) NPs, alkaline CometChip assay was performed. A549 cells were treated with 1 mg mL$^{-1}$ of AGuIX® or AGuIX-Bi (30Gd/70Bi) along with untreated controls for 24 h. CometChip preparation and cell loading were performed as described previously [67, 68]. In brief, 10,000 cells per 100 μL were loaded in each well of 96-well plates for 30 min at 37°C and 5% $CO_2$. After gravity settling, the chips were rinsed with 1X DPBS (Gibco) to wash excess cells. Chips were then overlaid with 1% (w/v) molten low-melting agarose (Gold Biotechnology) prepared in 1X DPBS and allowed to gelate for 2 min at room temperature. All chemical reagents used for CometChip assays were from Sigma or VWR unless otherwise stated. The chips were irradiated using a Small Animal Radiation Research Platform (SARRP, Xtrahl, Inc., Suwanee, GA) at 6 Gy or 10 Gy. A mock-irradiated chip was prepared and handled identically except without radiation exposure. Post irradiation, the chips were immediately submerged in ice-cold alkaline lysis solution (2.5 M NaCl, 100 mM $Na_2$EDTA, 10 mM Tris, pH 10 with 1% Triton X-100) and kept overnight at 4°C. Alkaline DNA unwinding was performed using alkaline buffer (0.3 M NaOH and 1 mM $Na_2$EDTA in distilled water) at 4°C for 40 min followed by electrophoresis in the same DNA unwinding buffer at 4°C for 30 min at 1V cm$^{-1}$ and a constant current of 300 mA. Chips were neutralized using 0.4 M Tris-HCl buffer (pH 7.5) and stained using 1X (0.01% (v/v) of 10,000X stock) SYBR gold nucleic acid gel stain (Invitrogen, cat. No. S11494) for 10 min at room temperature. Fluorescence comet images were captured at 10X magnification using an inverted LED fluorescence motorized microscope (Zen 3.2 pro blue edition, Zeiss Apotome 2, Carl Zeiss microscopy, GmbH). At least 500 cells were scored for each condition using Trevigen comet analysis software (R&D systems, MN, USA).

**Chemosensitivity assay**

Cells were seeded at a density of $0.1 \times 10^6$ cells mL$^{-1}$ in 96-well plates with 0.1 mL of culture medium. The cells were treated with the following conditions: 1 mg mL$^{-1}$ AGuIX®, 1 mg mL$^{-1}$ AGuIX-Bi (30Gd/70Bi), and radiation at 6 Gy or 10 Gy for 24 and 48 hours. After the designated treatment period, the cells were stained with Alexa Fluor 488-conjugated annexin V (AxV) and propidium iodine (PI) in 10X Annexin V binding buffer (100 mM HEPES, 40 mM KCl, 1.4 M NaCl, 7.5 mM MgCl2, 25 mM CaCl2, pH 7.4) at a dilution of 1:500. The staining solution was then added to the cells at a 1:10 dilution, and the cells were incubated on ice for 20 minutes. AxV/PI positivity was subsequently measured using an Attune NxT flow cytometer (Thermo Fisher).

**Internalization study**

Cells were seeded at a density of 50,000 cells on microscopy slides overnight. A549 cells were incubated with each Cy5.5 tagged NP formulation (1 mg mL$^{-1}$) overnight. Cells were fixed, blocked, and permeabilized with 10% FBS and 0.3% Triton X-100. Cells were stained with Cellmask Orange Plasma membrane stain (ThermoFisher) and then mounted with DAPI and Flouromount-G. Images were taken using a Zeiss AxioObserver microscope 63x.

**Animal tumor model and *in vivo* studies**

Animal studies were approved by Dana Farber Cancer Institute (DFCI) Institutional Animal Care and Use Committee (IACUC, Protocol Number 14-032) and conducted in full compliance with the Association for the Assessment and Accreditation of Laboratory Animal Care, governmental and institutional regulations and principals outlined in the United States Public Health Service Guide. Female 4 - 6 week athymic *Fox1$^{nu}$* mice were used (Charles River Laboratories, Wilmington, USA) weighing 16-20 g. Mice were fed ad libitum with standard food pellets and water. Mice were inoculated subcutaneously with 3 x 10$^6$ A549 cells in 100 µL$^{-1}$ into the dorsolateral right flank and allowed to grow for 2 - 3 weeks or until tumors reached 3 - 4 mm in diameter. Tumor volume was determined with calipers using the formula: $volume = \frac{length \times width^2}{2}$. Mice were given retroorbital, intravenous injections of 300 mg kg$^{-1}$ of either AGuIX® or AGuIX-Bi (30Gd/70Bi) followed by 10 Gy radiation either 30 mins or 24 h post-injection, respectively (n = 6 – 9). Control mice that did not receive nanoparticles were given saline. Mice were euthanized if tumor volume surpassed 10% of total animal body weight or if any other signs of distress were identified (ulcerations > 10 mm, lethargy, etc).

**ICP-MS biodistribution and pharmacokinetics**

ICP-MS measurements were performed to confirm the presence and quantify each organ's Gd and Bi content at different times after retro-orbital intravenous administration. Mice were injected retro-orbitally with 300 mg kg$^{-1}$ (~6 mg animal$^{-1}$) of AGuIX® or AGuIX-Bi (30Gd/70Bi) and sacrificed at 1 h, 4 h, 24 h, and 7 days (n = 3 animals timepoint$^{-1}$). Tumors, blood, and other vital organs (liver, kidney, spleen, lungs, and heart) were collected for inductively coupled plasma mass spectrometry (ICP-MS). Tumors and organs were dissected, weighed, and digested in HNO$_3$ (69%) through microwave Multiwave 5000 (Anton Paar, Austria) and analyzed by ICP-MS Nexion 2000 B (Perkin-Elmer).

**TDA-ICP-MS**

Experiments were performed on a P/ACE MDQ capillary electrophoresis instrument coupled to an ICP-MS (Agilent 7700) for Gd and Bi detection. Following a previously established protocol [55], TDA-ICP-MS experiments were conducted in 75 µm i.d. x 64 cm hydroxypropylcellulose modified capillaries. NP solutions (1 g L$^{-1}$) were prepared just before analysis and hydrodynamically injected (0.3 psi, 3 s). Samples were mobilized by applying a mobilization pressure of 0.7 psi using TRIS 10 mM, NaCl 125 mM, pH 7.4. Detection was performed by ICP-MS at m/z= 158 et 209.

Assuming populations of different sizes in the sample, (e.g. small molecules and nano-objects), the taylorgram obtained by Taylor Dispersion Analysis can be assimilated to the sum of several Gaussian signals. Deconvolution and integration of the peak were performed using the Origin 8.5 software to obtain molecular diffusion coefficients and then mean hydrodynamic diameters.

**SARRP radiation treatment planning and delivery**

A Small Animal Radiation Research Platform (SARRP, Xtrahl, Inc., Suwanee, GA) was used to perform image-guided radiation studies *in vitro* and *in vivo*. The DFCI SARRP therapy beam was initially measured with an ADCL-calibrated ion chamber, and the output was checked monthly. Mice were anesthetized with a 1 - 3% isoflurane mixture and an airflow of 1.5 L min$^{-1}$. A cone-beam computed tomography (CBCT) image was acquired at 60 kVp (0.8 mA current with a 1 mm Al filter) to identify the tumor isocenter in MuriSlice. A single posterior-anterior 220 kVp (13 mA current with a 0.15 mm Cu filter) beam delivered a dose of 10 Gy to the isocenter with a 1 x 1 cm$^2$ collimated field. Dosimetry studies were carried out in the MuriPlan treatment planning software to determine the radiation dose distribution in the tumor (**Figure S27**).

**Toxicology**

Mice bearing A549 subcutaneous tumors were treated with either saline or 300 mg kg$^{-1}$ AGuIX-Bi (30Gd/70Bi) followed by 10 Gy irradiation 24h post-injection. Tumors and other vital organs (liver, lungs, spleen, kidney, and heart) were collected either 24 h or 1 mo post-therapy and histologically stained (H&E). Liver and kidney sections were additionally stained for caspase-3 to evaluate damage. A pathologist assessed each sample to identify any short- or long-term toxicity effects. Blood was collected and allowed to clot before collection of serum. Serum was analyzed for kidney (BUN), liver (ALT, AST, GGT), and systemic inflammatory damage (LDH) at the Beth Israel Deaconess Medical Preclinical Murine Pharma Core.

**Magnetic resonance imaging**

MRI experiments were conducted on a Bruker BioSpec 7T/30 cm USR horizontal bore Superconducting Magnet System (Bruker Corp., Billerica, MA) outfitted with the B-GA12S2 gradient and integrated with up to a 2$^{nd}$ order room temperature shim system that provides a slew rate of 3440 T m$^{-1}$ s$^{-1}$ and a maximum gradient amplitude of 440 mT m$^{-1}$ and. The Bruker 35 mm ID birdcage volume radiofrequency (RF) coil was utilized for RF excitation and receiving. Experiments were performed with animals anesthetized at 1.5% isoflurane mixed in medical air and a flow rate of about 2 L min$^{-1}$. Body temperature was maintained using a warm air fan at 37°C. Animal respiration and temperature were regulated by the 1025T SAII (Sa Instruments Inc., Stony Brook, NY) monitoring and gating system. A pressure-transducer for respiratory gating was placed on the abdomen. Bruker Paravision 6.0.1 was used for MRI data acquisition, T1 data analysis, and T1 map generation. T1 weighted tumor images were obtained using a fast spin echo (RARE) sequence with fat suppression using the following parameters: TR = 650 ms, TE = 16 ms, FOV = 30 x 30 mm$^2$, matrix size = 128 x 128, spatial resolution = 156 x 156 μm$^2$, number of slices = 16, thickness = 1 mm, rare factor = 3, number of averages = 5, acquisition time = 5:00 mins in the axial view. In the coronal view, parameters were the same except for matrix size = 192 x 192, spatial resolution = 156 x 156 μm$^2$, number of slices = 20, and slice thickness = 0.5 mm. T1/T2 map sequence utilized a spin echo readout, axial view, 6 T1 acquisitions with TR = 7000, 5500, 3000, 1500, 1245, 487 ms, 5 T2 acquisitions with echo spacing (ESP) = 6.3, 18.9, 31.45, 44, 56.6 ms, number of signal averages (NSA) = 1, flip angle (FA) = 90/180, echo train length (ETL) = 2, bandwidth (BW) = 50 MHz, matrix size = 128 x 128, FOV = 30 x 30 mm$^2$, thickness = 1 mm, scan time = 15:22 mins. Only data related to the T1 map were analyzed for phantoms of AGuIX® and AGuIX-Bi (30Gd/70Bi) (**Figure S28**) to determine NP relaxivity at 7 T. Reconstructed MR images of mice were bias field corrected and normalized to surrounding muscle (MIM software) as no NP accumulation is expected within healthy muscle tissue.

**Statistical analysis**

Results are reported as the mean ± standard error of the mean (SEM) as indicated in figure legends. All statistical analysis was performed with GraphPad Prism (version 8). A Log-rank Mantel-Cox test with the Bonferroni correction for multiple comparisons was used to assess median tumor-doubling time. A two-way ANOVA with Tukey's multiple comparisons test was used to assess tumor growth and alkaline CometChip assay results. A one-way ANOVA was used to compare three or more means with Kruskal-Wallis nonparametric test (non-Gaussian population) for mean tumor doubling time and γH2AX *in vitro* assay. All *in vitro* studies were conducted in triplicates. Differences of $p < 0.05$ were considered statistically significant.

**Supplementary Material**

Supplementary Material is available online at Theranostics or from the author.

**Acknowledgments**

We would like to thank Dr. Thomas Brichart for his help in creating the nanoparticle illustration through the Blender software, and his advice throughout the project. The research reported in this publication was partially funded by the CNRS (French National Centre for Scientific Research) IRP Radioboost. Research reported in this publication was supported by the National Cancer Institute of the National Institute of Health under award number R01CA240804. The research reported in this publication was partially supported by NCI award number 5R25CA174650. Research reported in this publication was partially supported by pilot project grant from Harvard-NIEHS Center for Environmental Health (P30ES000002). We thank Dr. Bevin Engelward for sharing some of the materials needed to perform the CometChip assay. The content is solely the responsibility of the authors and does not necessarily represent the official views of the National Institute of Health.

**Conflicts of Interest**

François Lux and Olivier Tillement have to disclose the patent WO2011/135101. François Lux, Olivier Tillement, and Ross Berbeco have to disclose the patent US62431607. Paul Rocchi, Léna Carmès Tristan Doussineau, and Sandrine Dufort are employees of NH TherAguix which develops AGuIX® nanoparticles. François Lux, Olivier Tillement, Tristan Doussineau, and Sandrine Dufort possess shares in the company, NH TherAguix.

**Author Contributions**

The work presented here was performed in collaboration among all authors. Needa Brown designed the methods and experiments, performed the laboratory studies, and analyzed the data related to in vitro and in vivo studies. Paul Rocchi designed the methods and experiments, performed the laboratory studies, and analyzed the data related to nanoparticle formulation and characterization. Needa Brown and Paul Rocchi wrote the manuscript as co-first authors. Léna Carmès helped conduct ICP-MS experiments. Meghna Iyer and Léa Seban helped conduct in vitro experiments and histology slide imaging. Johany Penailillo and Ruben Carrasco helped with histology slide imaging and analysis. Romy Guthier, Meghna Iyer, Léa Seban, Toby Morris, Stephanie Bennett, Michael Lavelle, Chris Williams, Elizabeth Huynh, Zhaohui Han, and Evangelia Kaza helped conduct in vivo experiments. Tristan Doussineau helped with nanoparticle formulation and characterization. Guillaume Bort helped with nanoparticle formulation and characterization and revised the manuscript critically. Sneh M. Toprani conducted, and Zachary D. Nagel supervised the alkaline CometChip experiments, imaging, and scoring. Xingping Qin conducted, and

Supplementary Material

**Tuning ultrasmall theranostic nanoparticles for MRI contrast and radiation dose amplification**

*Needa Brown[#*], Paul Rocchi[#], Léna Carmès, Romy Guthier, Meghna Iyer, Léa Seban, Toby Morris, Stephanie Bennett, Michael Lavelle, Johany Penaililo, Ruben Carrasco, Chris Williams, Elizabeth Huynh, Zhaohui Han, Evangelia Kaza, Tristan Doussineau, Sneh M. Toprani, Xingping Qin, Zachary D. Nagel, Kristopher A. Sarosiek, Agnès Hagège, Sandrine Dufort, Guillaume Bort, François Lux, Olivier Tillement, and Ross Berbeco[*]*

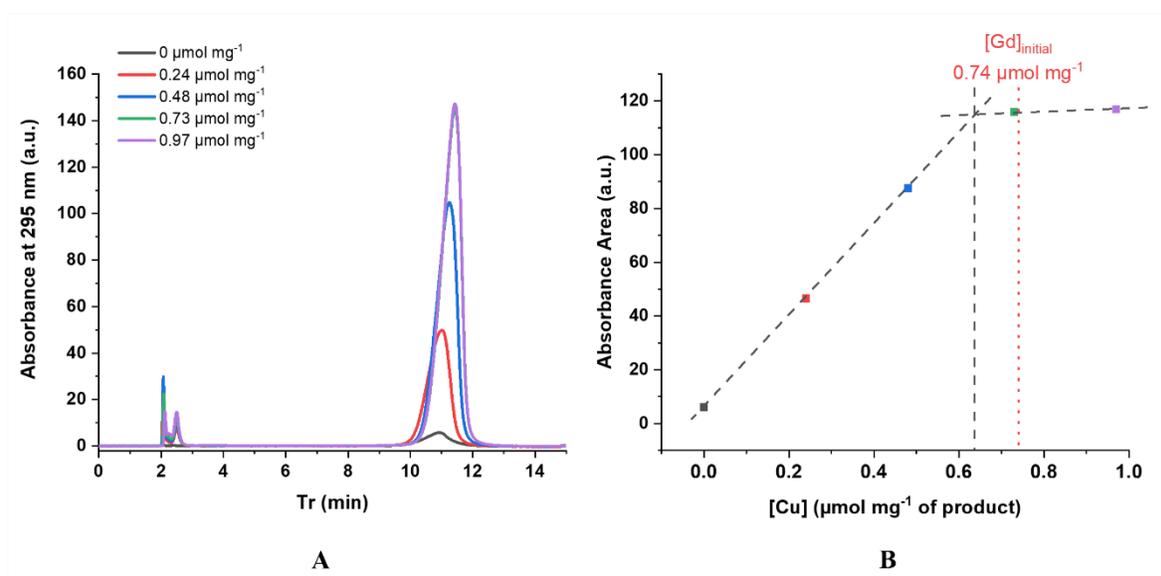

**Figure S1:** Measurement of free DOTAGA at the beginning of the bismuth chelation. A) Chromatograms of samples with increasing amount of $Cu^{2+}$. The increase in absorbance at 295 nm is due to the formation of $DOTAGA@(Cu^{2+})$. B) Measured area depends on the amount of $Cu^{2+}$. The slope break shows the amount of free DOTAGA in the product.

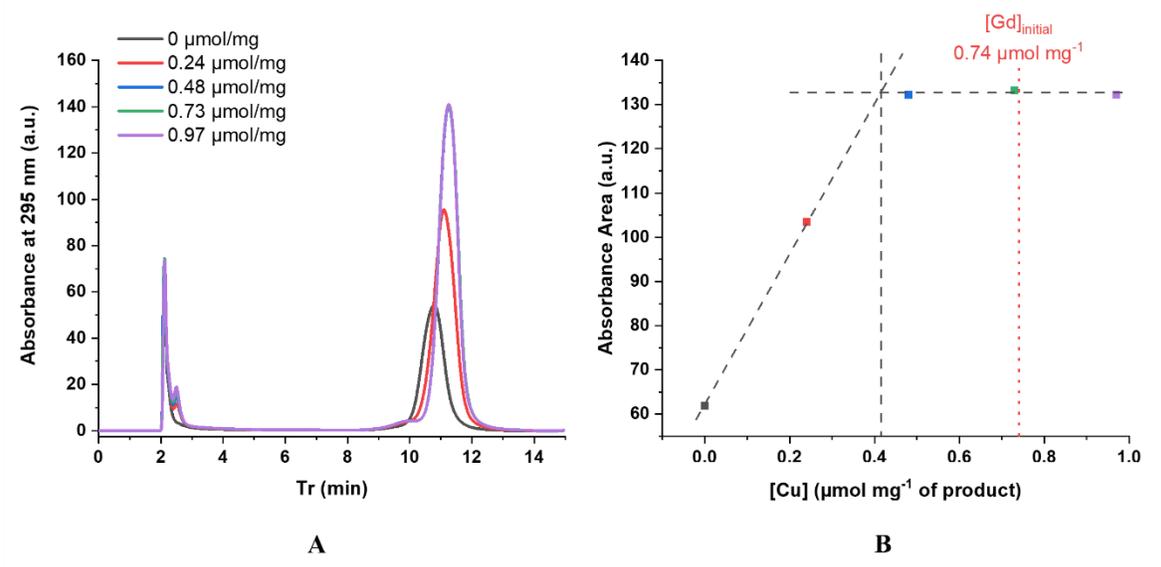

**Figure S2:** Measurement of free DOTAGA after the first chelation bismuth chelation of 30%. A) Chromatograms of samples with increasing amount of $Cu^{2+}$. The increase in absorbance at 295 nm is due to the formation of DOTAGA@($Cu^{2+}$). B) Measured area depends on the amount of $Cu^{2+}$. The slope break shows the amount of free DOTAGA in the product.

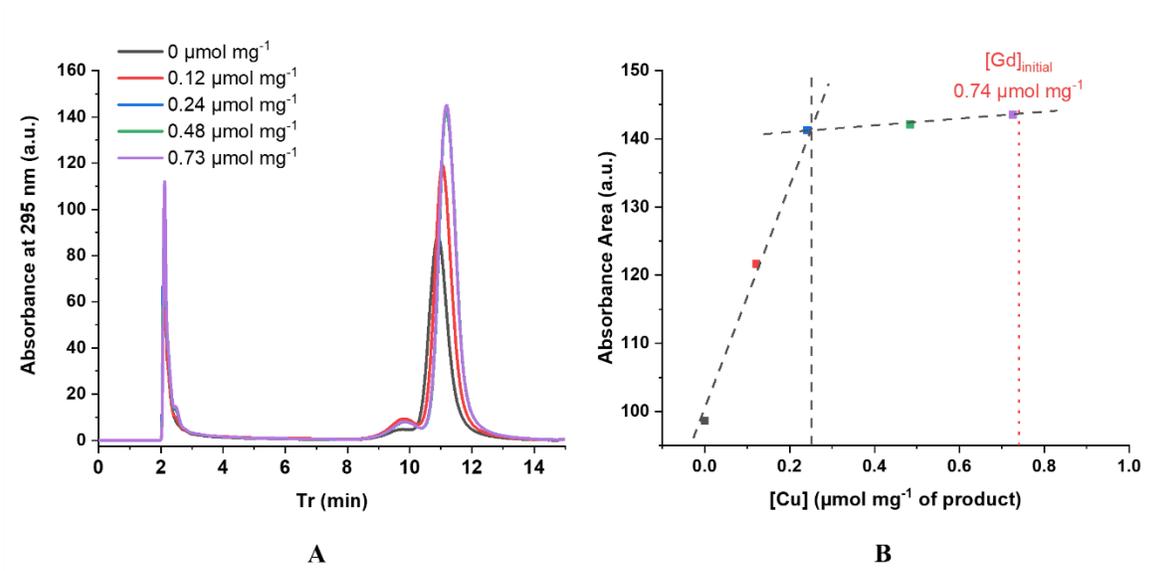

**Figure S3:** Measurement of free DOTAGA after the second chelation bismuth step of 20%. A) Chromatograms of samples with increasing amount of $Cu^{2+}$. The increase in absorbance at 295 nm is due to the formation of DOTAGA@($Cu^{2+}$). B) Measured area depends on the amount of $Cu^{2+}$. The slope break shows the amount of free DOTAGA in the product.

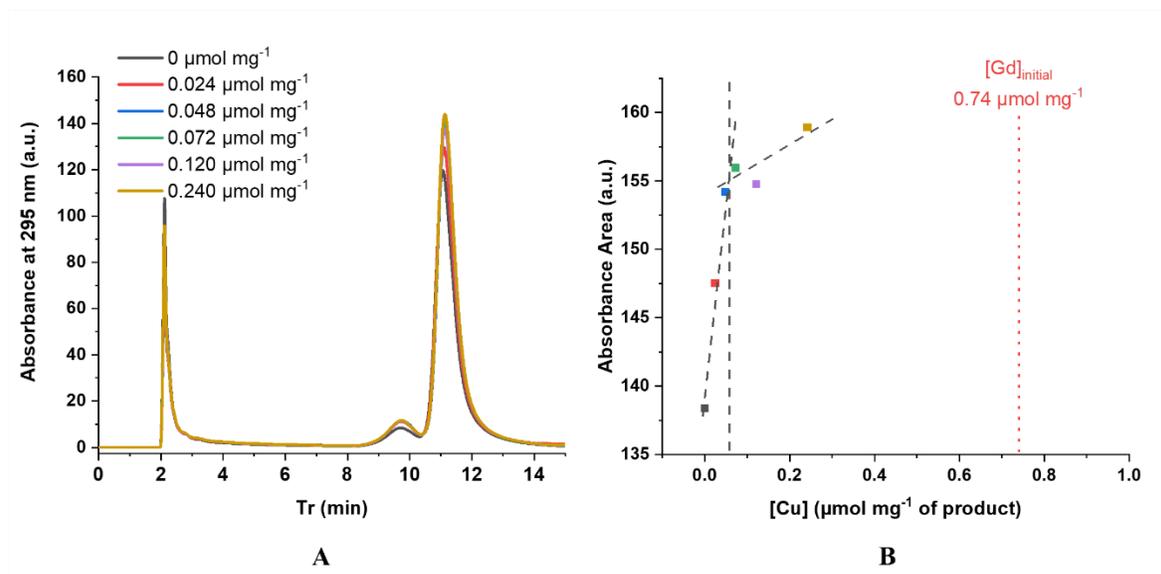

**Figure S4:** Measurement of free DOTAGA after the third chelation bismuth step of 20%. A) Chromatograms of samples with increasing amount of $Cu^{2+}$. The increase in absorbance at 295 nm is due to the formation of DOTAGA@($Cu^{2+}$). B) Measured area depends on the amount of $Cu^{2+}$. The slope break shows the amount of free DOTAGA in the product.

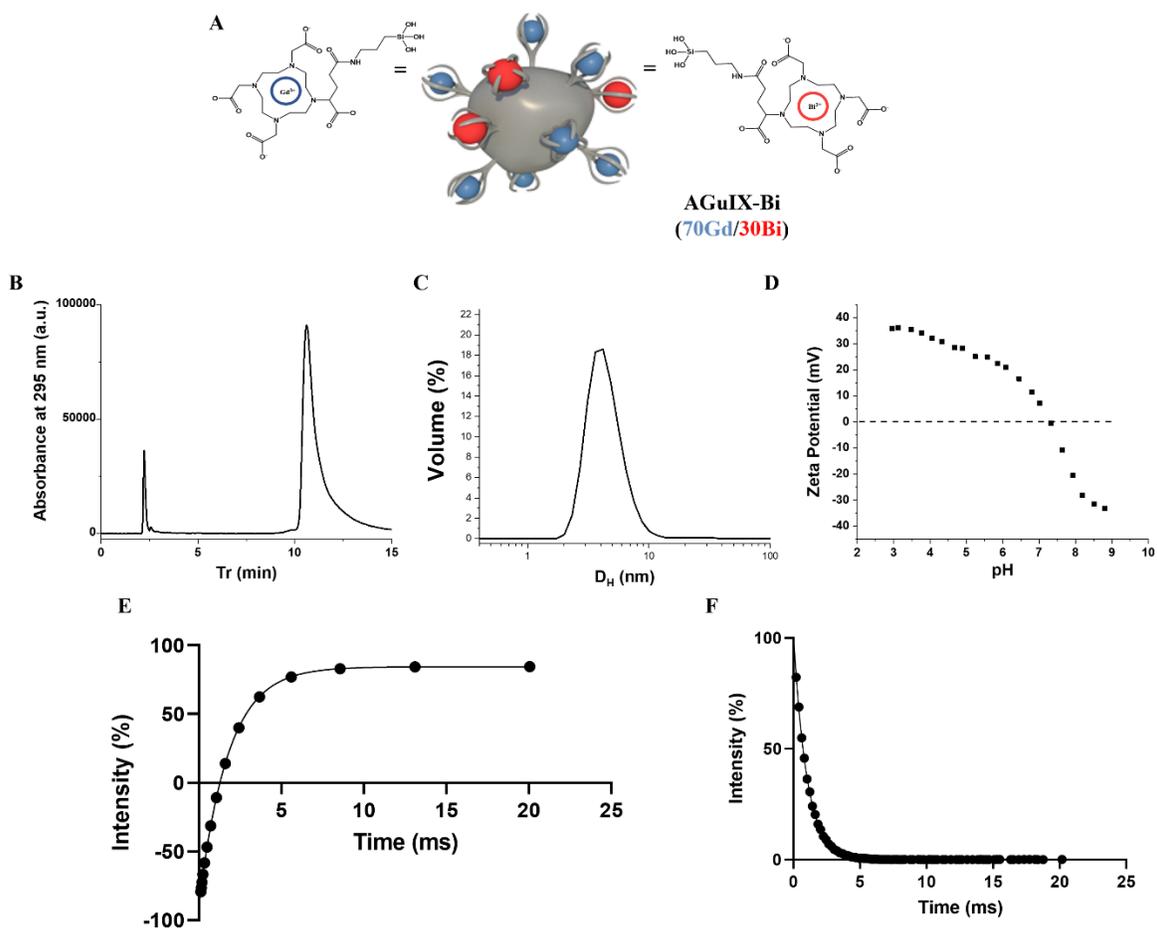

**Figure S5:** A) Schematic representation of AGuIX-Bi (70Gd/30Bi) NP with detailed structure of chelated DOTAGA groups. B) HPLC-UV chromatograms of AGuIX-Bi (70Gd/30Bi) NP (10 µL, 10 g L$^{-1}$) recorded at 295 nm. C) Hydrodynamic diameter distribution in volume obtained by dynamic light scattering. D) Zeta potential vs pH for AGuIX-Bi (70Gd/30Bi) NP. Relaxivity measurements for E) $r_1$ and F) $r_2$ at 1.5 T of AGuIX-Bi (70Gd/30Bi) NP.

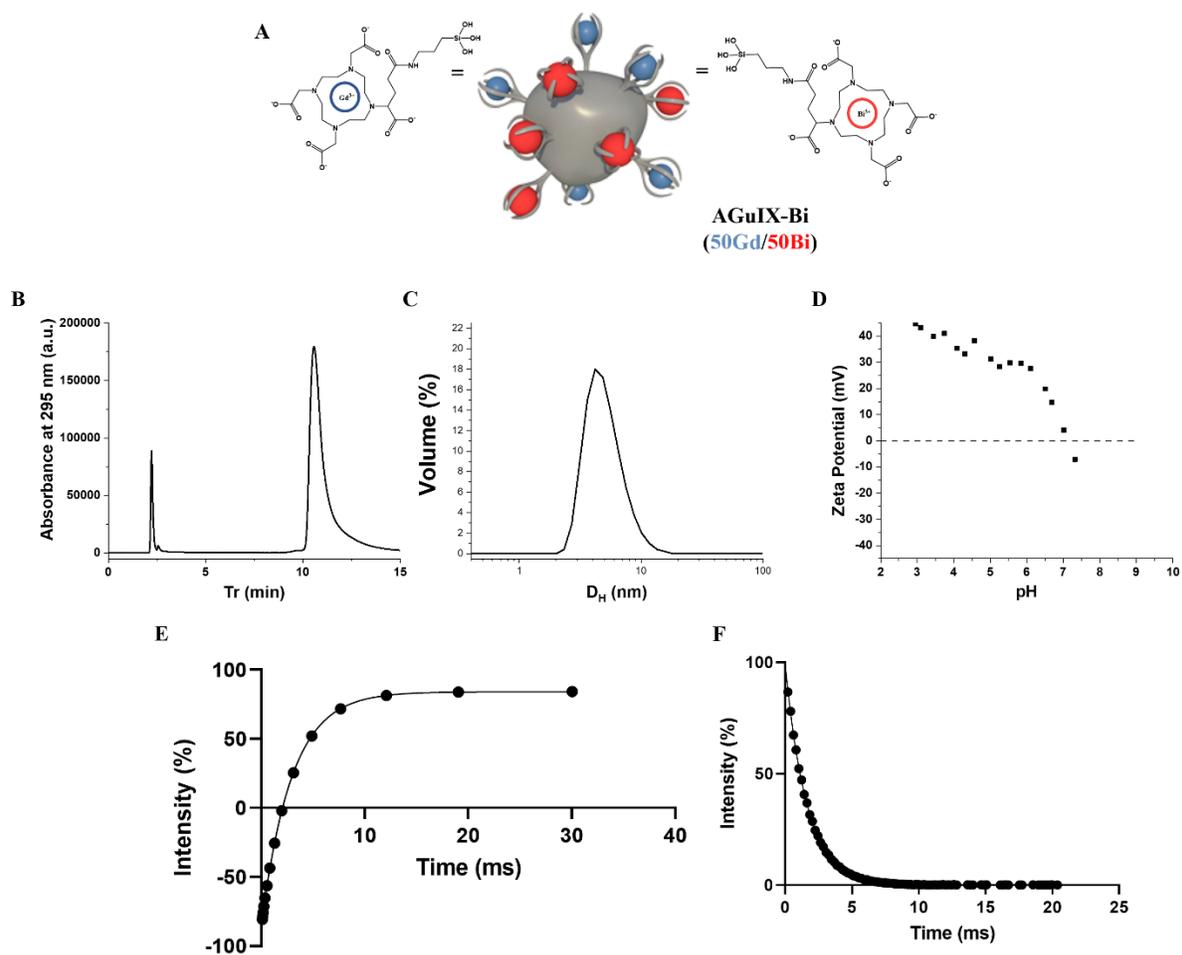

**Figure S6:** A) Schematic representation of AGuIX-Bi (50Gd/50Bi) NP with detailed structure of chelated DOTAGA groups. B) HPLC-UV chromatograms of AGuIX-Bi (50Gd/50Bi) NP (10 μL, 10 g L$^{-1}$) recorded at 295 nm. C) Hydrodynamic diameter distribution in volume obtained by dynamic light scattering. D) Zeta potential vs pH for AGuIX-Bi (50Gd/50Bi) NP. Relaxivity measurements for E) $r_1$ and F) $r_2$ at 1.5 T of AGuIX-Bi (50Gd/50Bi) NP.

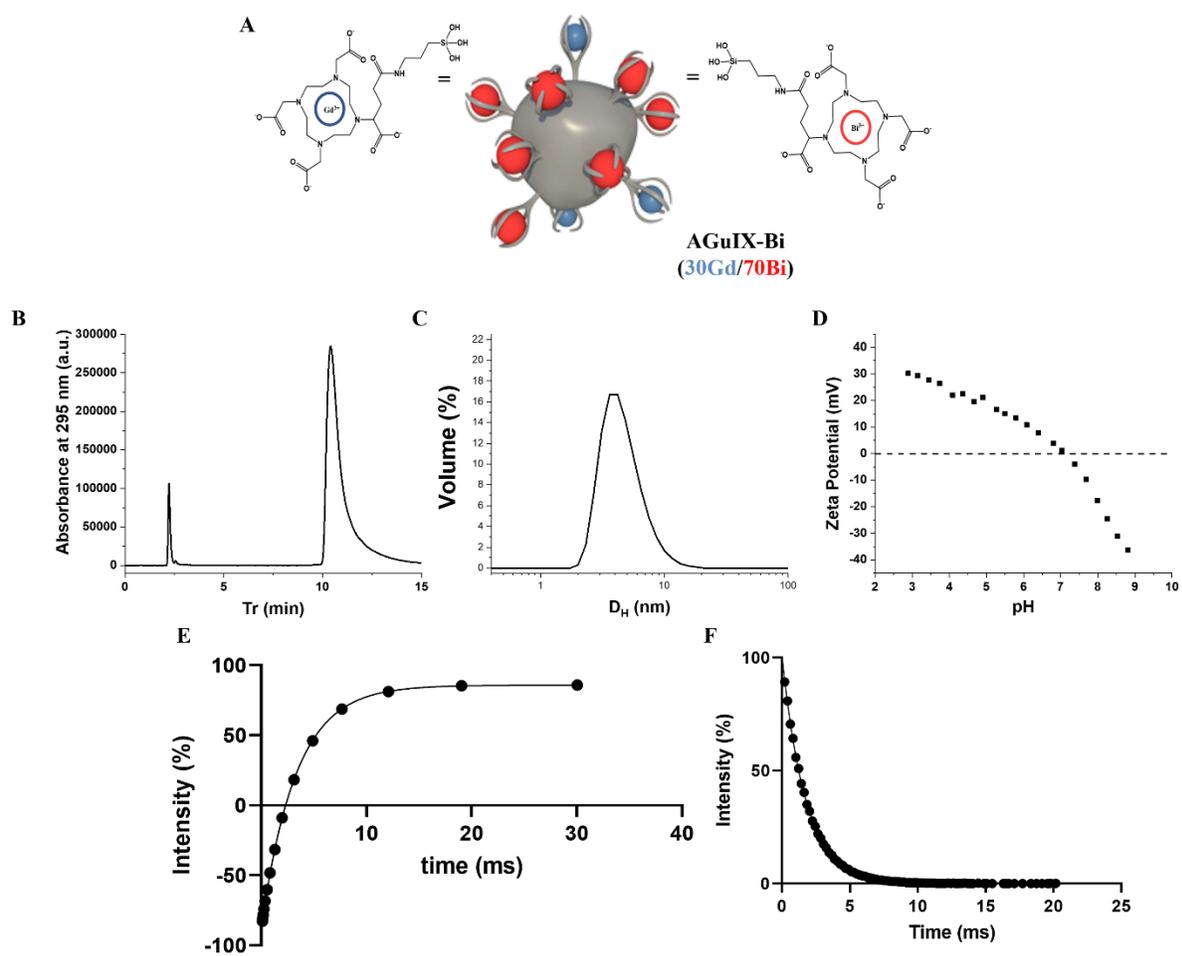

**Figure S7:** A) Schematic representation of AGuIX-Bi (30Gd/70Bi) NP with detailed structure of chelated DOTAGA groups. B) HPLC-UV chromatograms of AGuIX-Bi (30Gd/70Bi) NP (10 µL, 10 g L$^{-1}$) recorded at 295 nm. C) Hydrodynamic diameter distribution in volume obtained by dynamic light scattering. D) Zeta potential vs pH for AGuIX-Bi (30Gd/70Bi) NP. Relaxivity measurements for E) $r_1$ and F) $r_2$ at 1.5 T of AGuIX-Bi (30Gd/70Bi) NP.

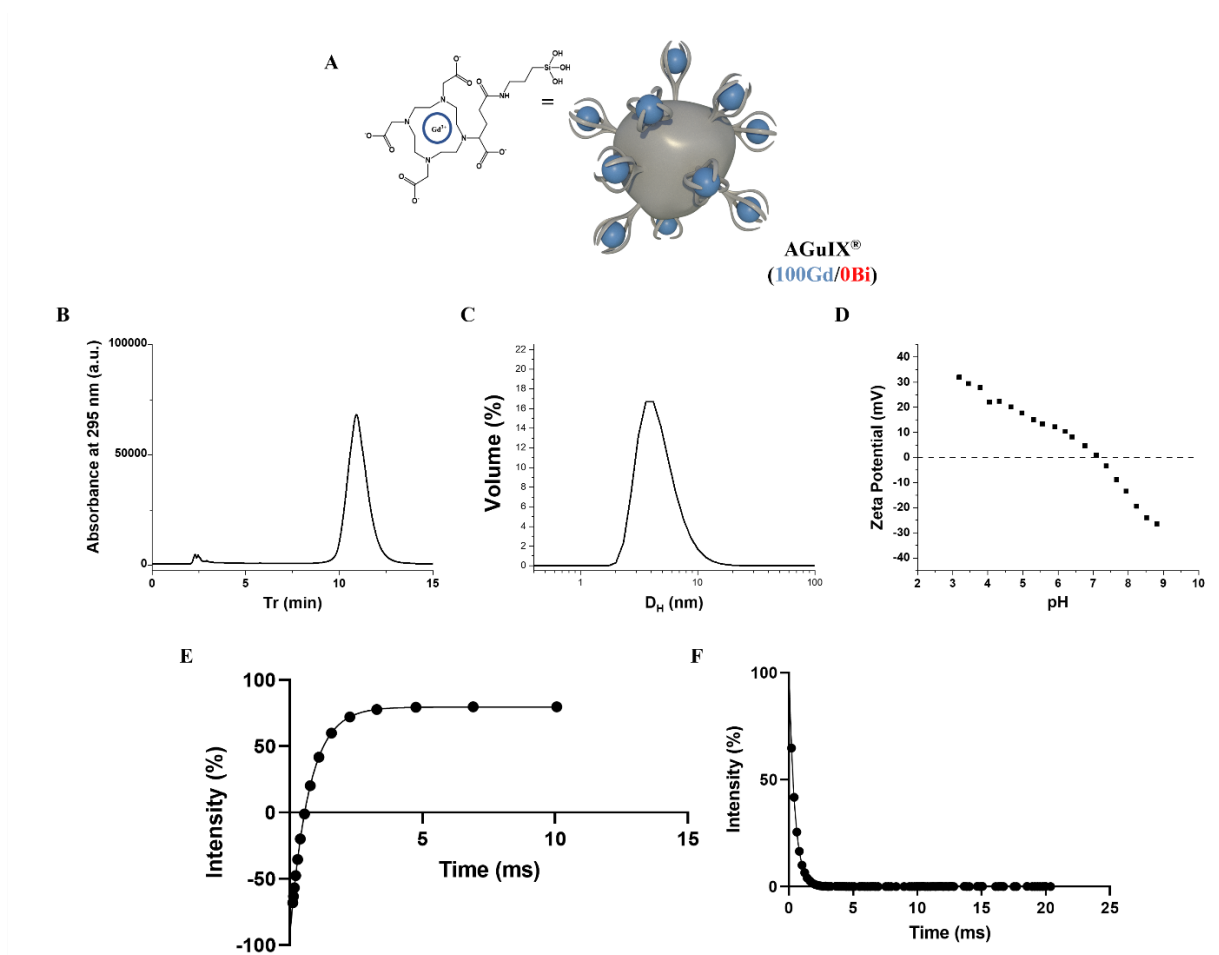

**Figure S8:** A) Schematic representation of AGuIX® NP with detailed structure of chelated DOTAGA groups. B) HPLC-UV chromatograms AGuIX® NP (10 µL, 100 g L$^{-1}$) recorded at 295 nm. C) Hydrodynamic diameter distribution in volume obtained by dynamic light scattering. D) Zeta potential vs pH for AGuIX® NP. Relaxivity measurements for E) $r_1$ and F) $r_2$ at 1.5 T of AGuIX® NP.

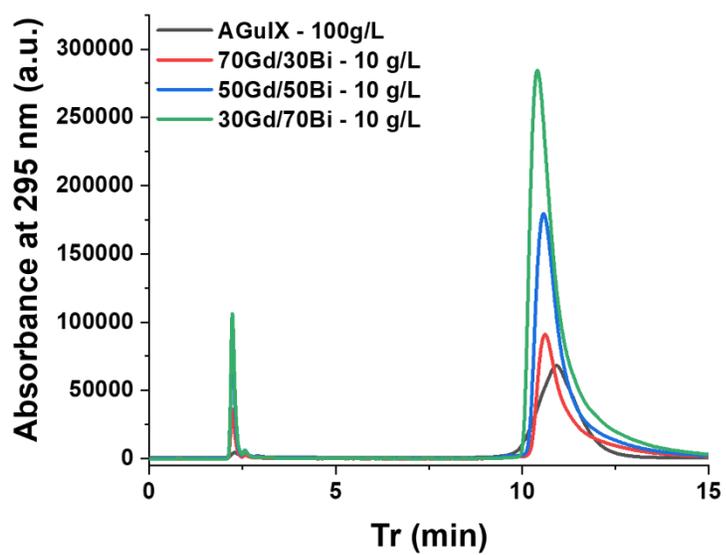

**Figure S9:** HPLC-UV chromatograms of AGuIX® particle (10 µL, 100 g L$^{-1}$) and the three final Gd/Bi (10 µL, 10 g L$^{-1}$) recorded at 295 nm. The increase of the absorbance at the level Tr is due to the creation of DOTAGA(Bi).

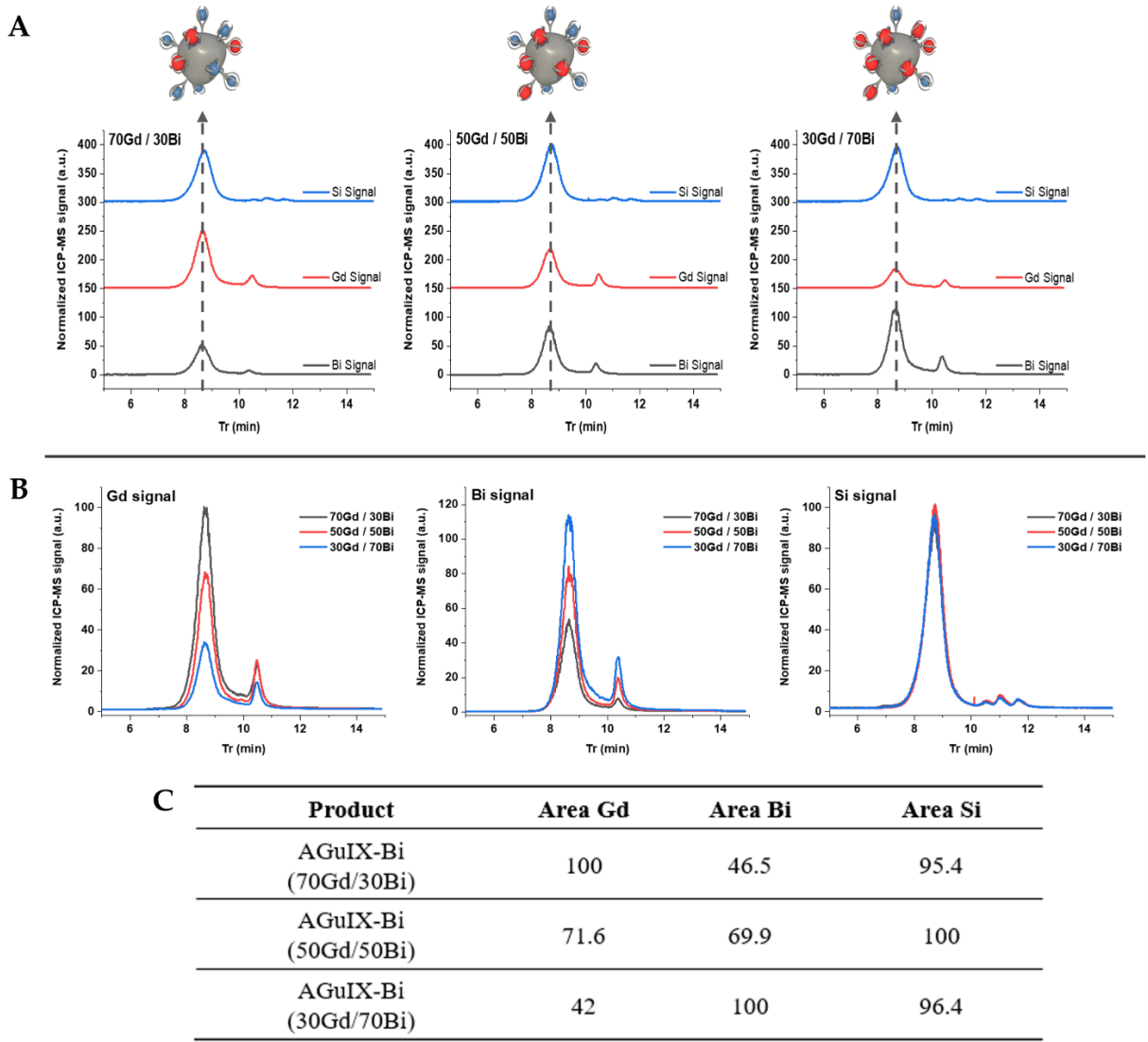

**Figure S10:** A) HPLC-SEC-ICP-MS chromatograms of the three final Gd/Bi following Gd, Bi and Si signals. We can observe the elemental ratio evolution between each product. B) HPLC-SEC-ICP-MS chromatograms of the three final Gd/Bi regrouped by specific element C) showing the ratio evolution of each measured element.

|  | **AGuIX®** | **20% Gd 0% Bi** | **20% Gd 30% Bi** | **20% Gd 70% Bi** |
|---|---|---|---|---|
| **Isoelectric point** | 7.15 | 5.58 | 6.63 | 6.95 |

**Figure S11:** IEP evolution all along the process. The decrease from one step to another is related to the generation of DOTAGA groups at the surface. In the contrary, the increase of the IEP is related to the chelation of a free DOTAGA by $Gd^{3+}$ or $Bi^{3+}$.

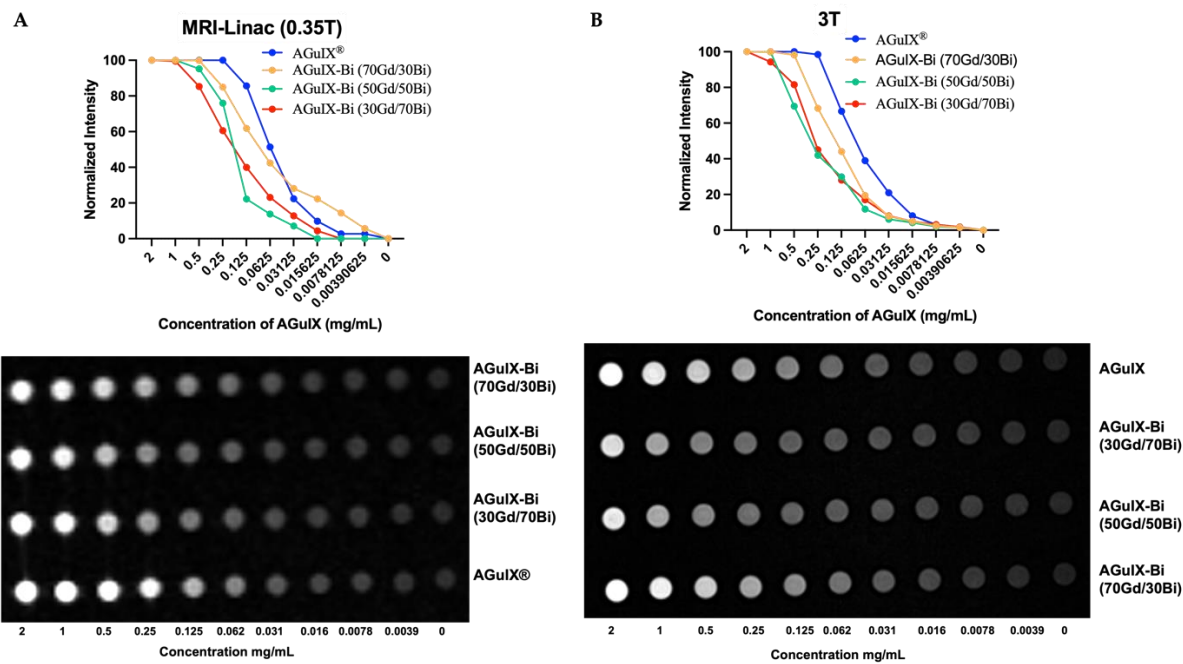

**Figure S12:** Agar phantom MRI images and signal intensity quantification of AGuIX® and various ratios of AGuIX-Bi nanoparticles in a A) 0.35T MRIdian MRI-Linac and B) 3T Siemens MR scanner. Signal intensity was normalized to highest concertation of nanoparticle.

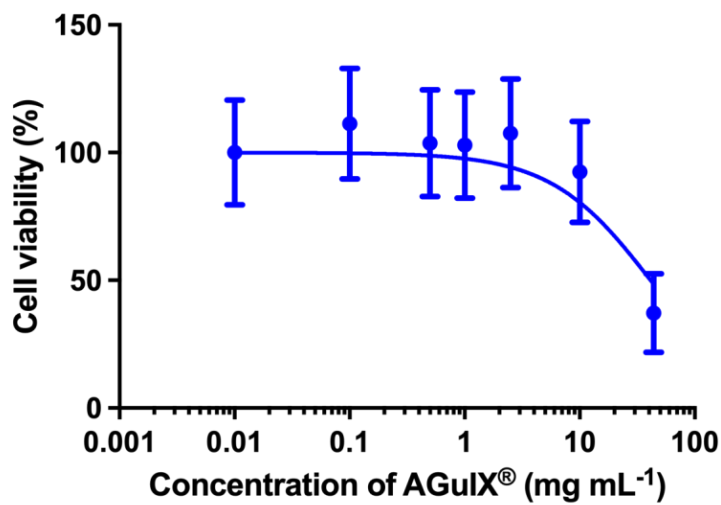

**Figure S13:** *In vitro* cytotoxicity of AGuIX® as a function of concentration 24 h post-incubation in a NSCLC A549 murine model (n = 3).

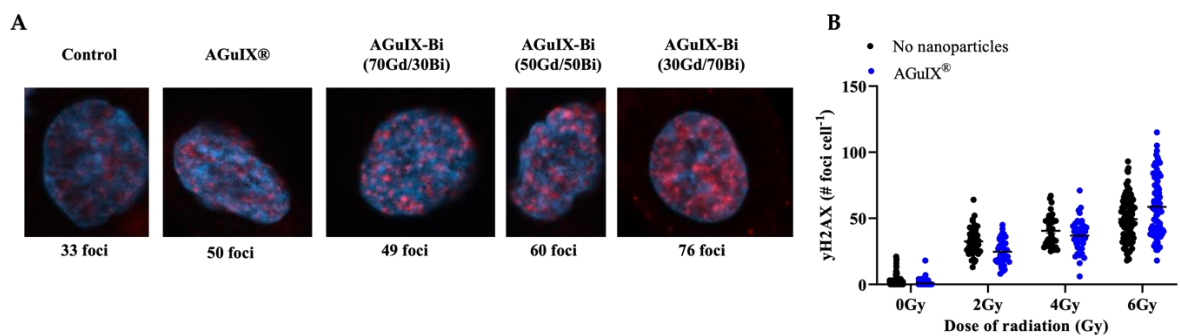

**Figure S14:** A) Representative single cell nuclei (blue) showing increasing γH2AX foci (red dots) with increasing ratio of Bi. B) Breaks in dsDNA following radiation was assayed using γH2AX for AGuIX® (n = 60-80 cells).

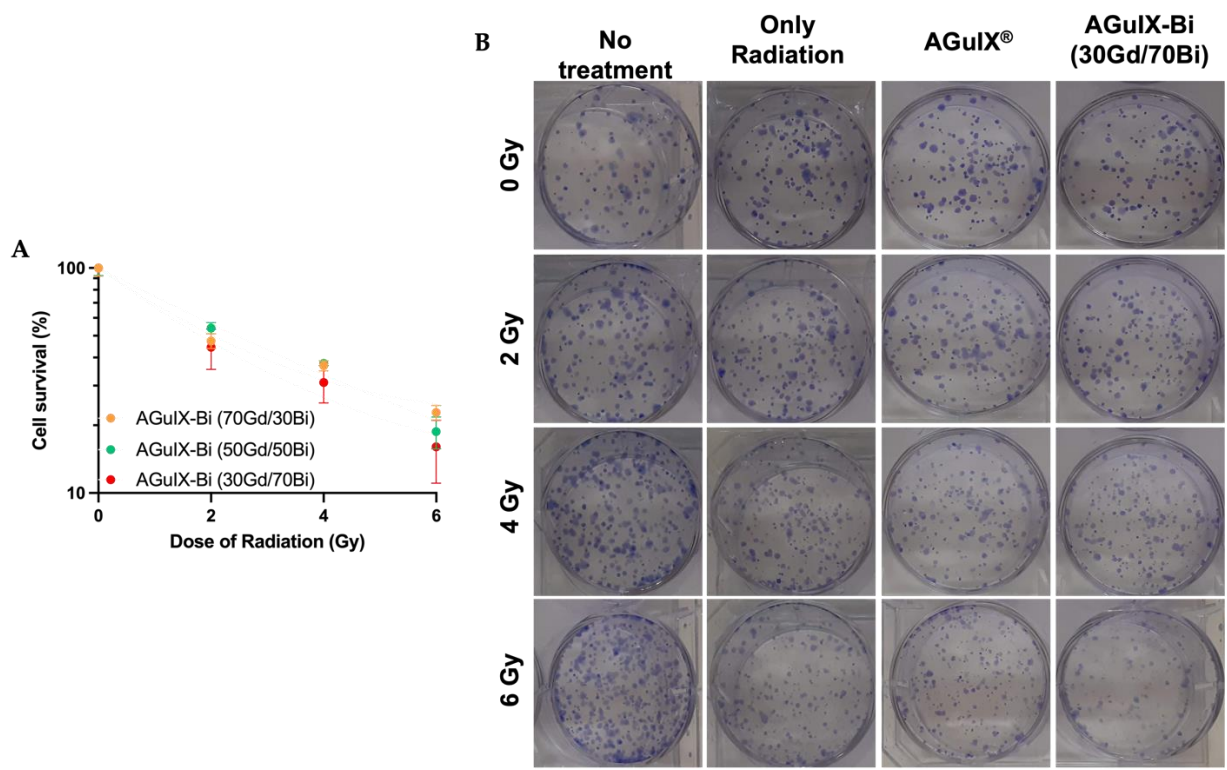

**Figure S15:** A) Clonogenic assay comparing the impact of increasing Bi ratio on cell survival. B) Clonogenic assay representative images.

F

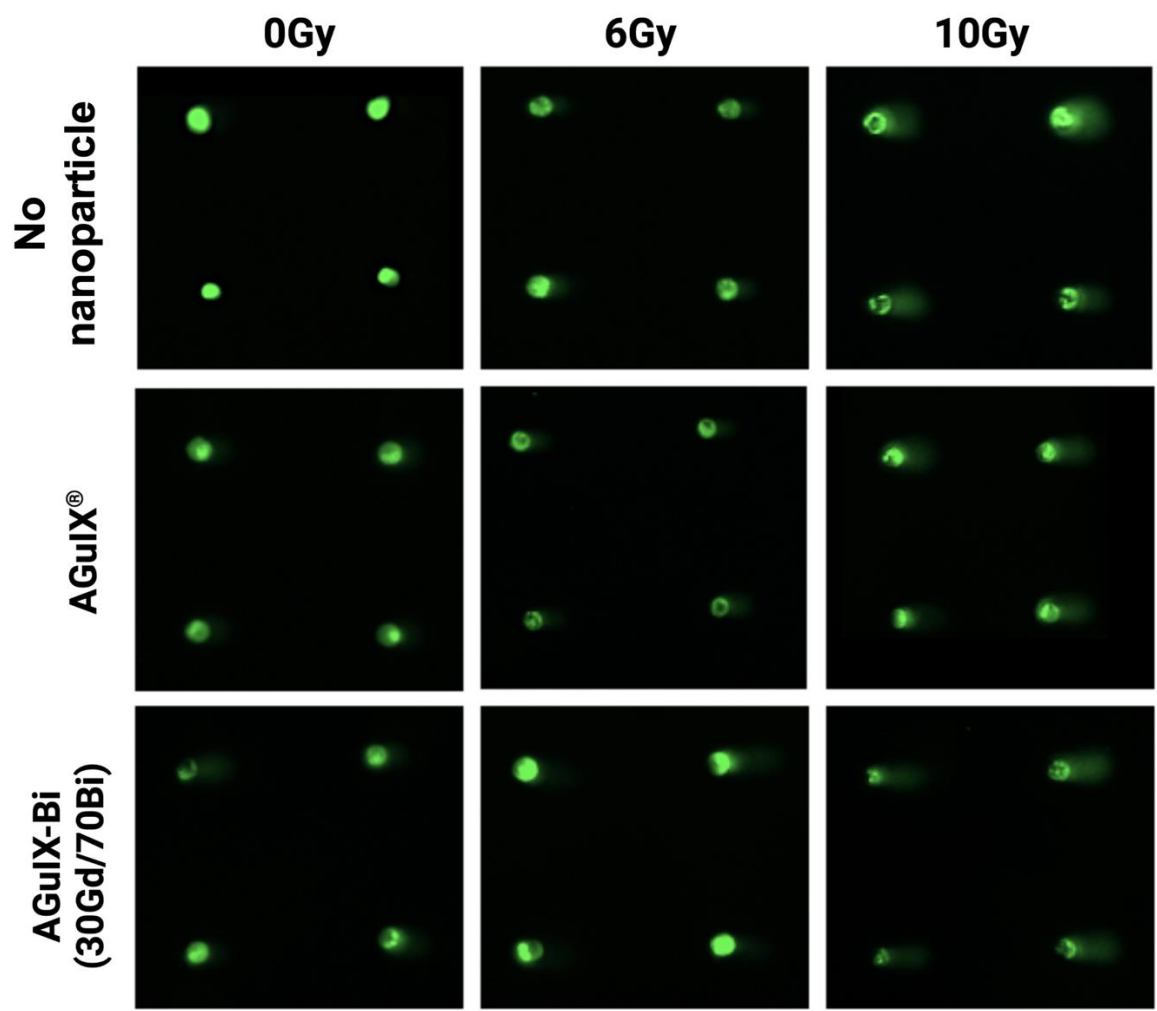

**Figure S16:** Representative comets from A549 cells treated with AGuIX® or AGuIX-Bi (30Gd/70Bi) with sham controls or irradiated at 6 Gy or 10 Gy.

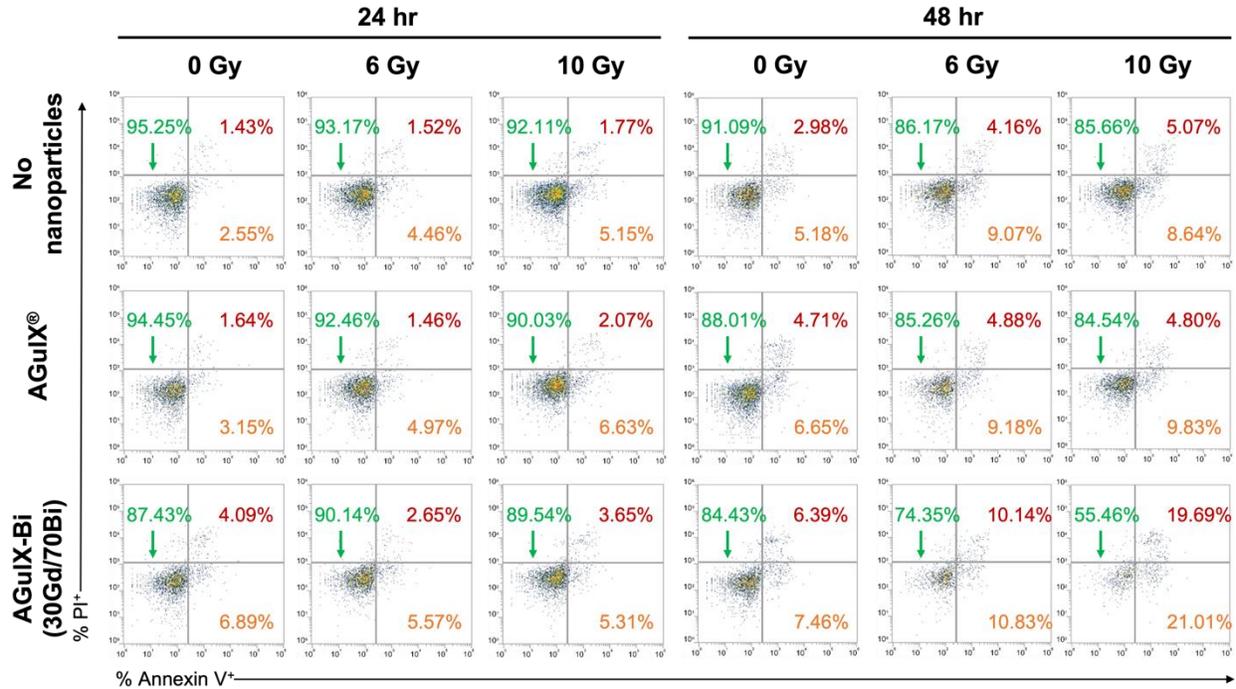

**Figure S17:** Representative flow cytometry data of cells positive for annexin V (AxV, orange), propidium iodine (PI, green), or AxV+PI+ (red). A549 cells treated with AGuIX® or AGuIX-Bi (30Gd/70Bi) with sham controls or irradiated at 6 Gy or 10 Gy.

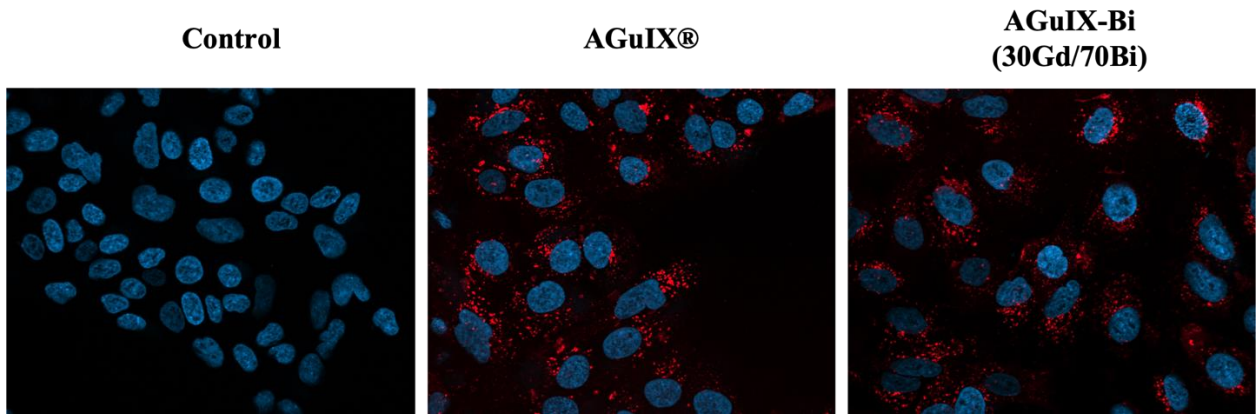

**Figure S18:** Representative nanoparticle (red) internalization images of A549 cells (blue) treated with either AGuIX® or AGuIX-Bi (30Gd/70Bi) for 24h.

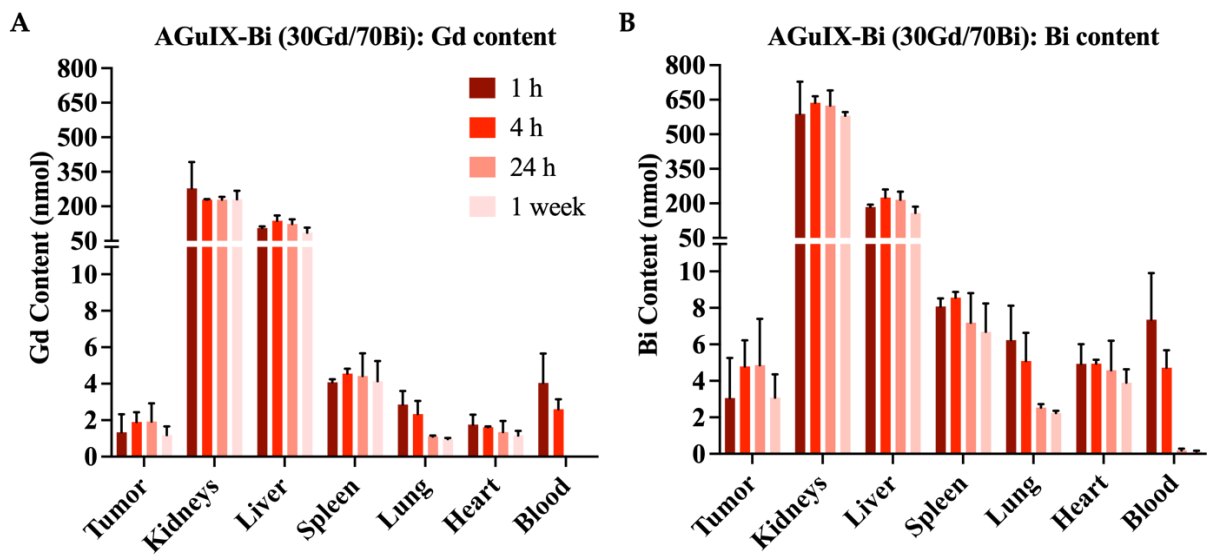

**Figure S19:** Raw ICP-MS nanomolar amount of A) gadolinium and B) bismuth in extracted tissue from mice treated with AGuIX-Bi (30Gd/70Bi). An average Bi/Gd molar ratio of 2.3 ± 0.56 was observed across all extracted tissue.

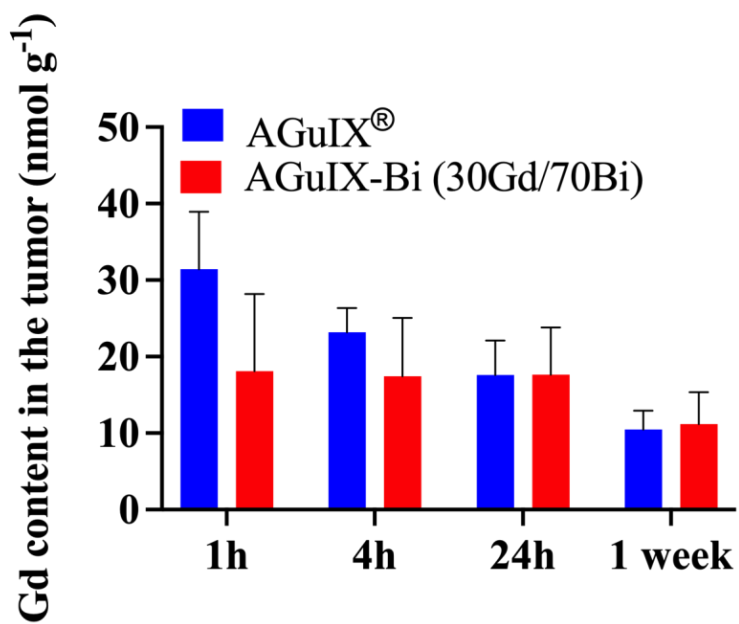

**Figure S20:** Raw ICP-MS nanomolar amount of gadolinium in extracted tumor tissue.

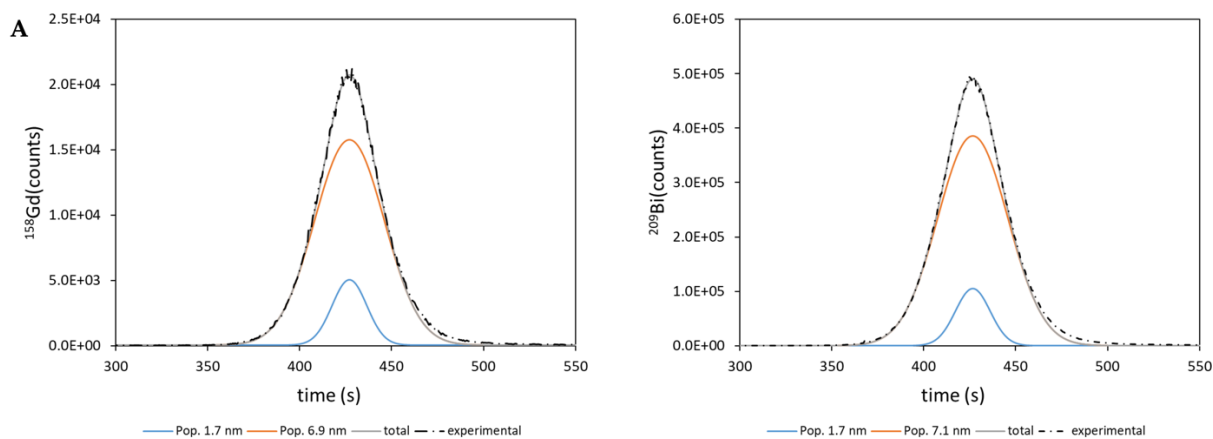

| Signal | Pop 1 (nm) | Metal proportion in Pop 1 (%) | Pop 2 (nm) | Metal proportion in Pop 2 (%) |
|---|---|---|---|---|
| $^{158}$Gd | 7.1 ± 0.3 | 86 ± 1 | 1.7 ± 0.1 | 14 ± 1 |
| $^{209}$Bi | 7.1 ± 0.5 | 87 ± 2 | 1.8 ± 0.4 | 13 ± 2 |

**Figure S21.** A) Representative taylorgram and corresponding bimodal fit for gadolinium (left) and bismuth (right) signal for AGuIX-Bi (30Gd/70Bi) nanoparticles. The signal (dashed black line) was fitted with the sum (solid gray line) of two gaussian curves (solid blue and orange lines) of the two populations within the sample. B) Average hydrodynamic diameters and Gd and Bi repartition from the TDA experiments (n = 3).

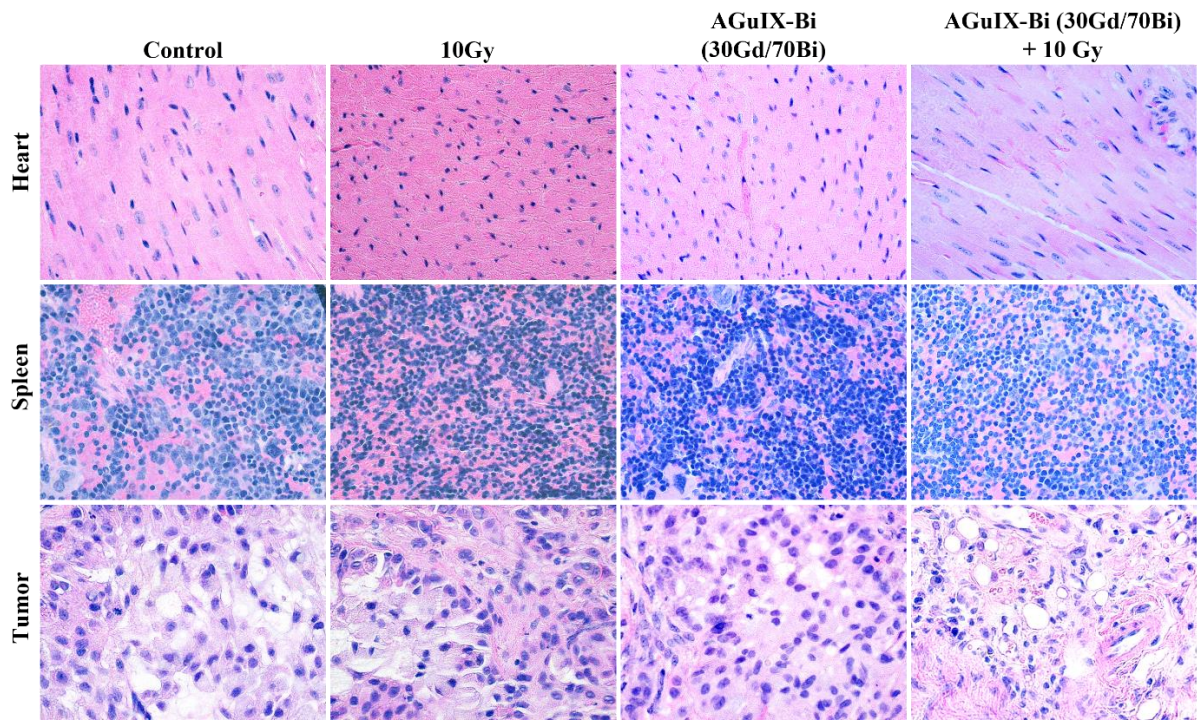

**Figure S22:** Histopathology of mice treated with either saline or AGuIX-Bi (30Gd/70Bi) and/or 10 Gy radiation 24 h post-treatment. No short-term toxicity was identified in heart, spleen, or tumor.

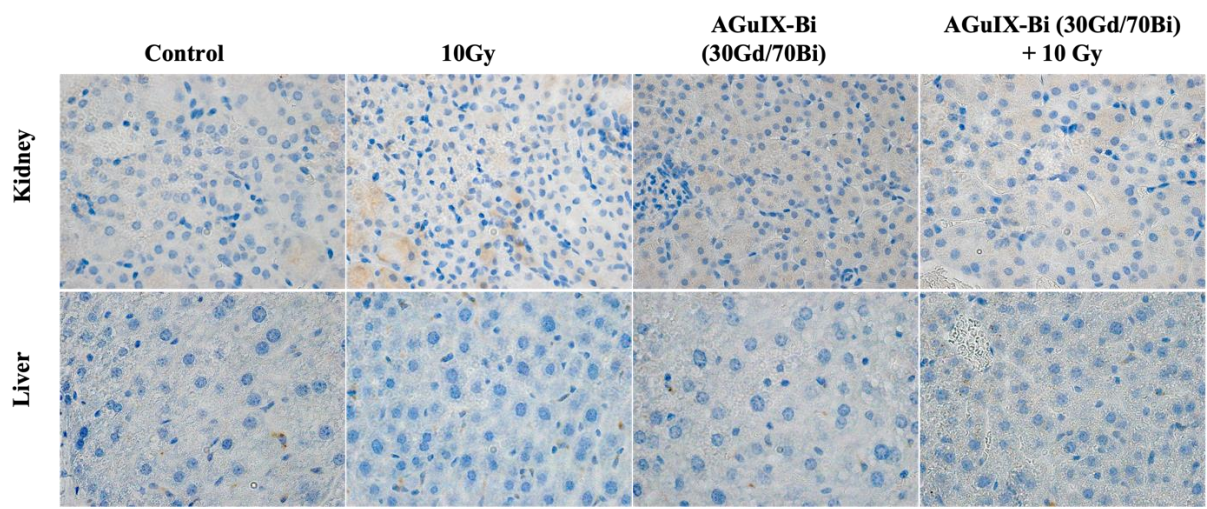

**Figure S23:** Caspase-3 staining of liver and kidney, which have the highest accumulation of nanoparticles, show no difference in staining.

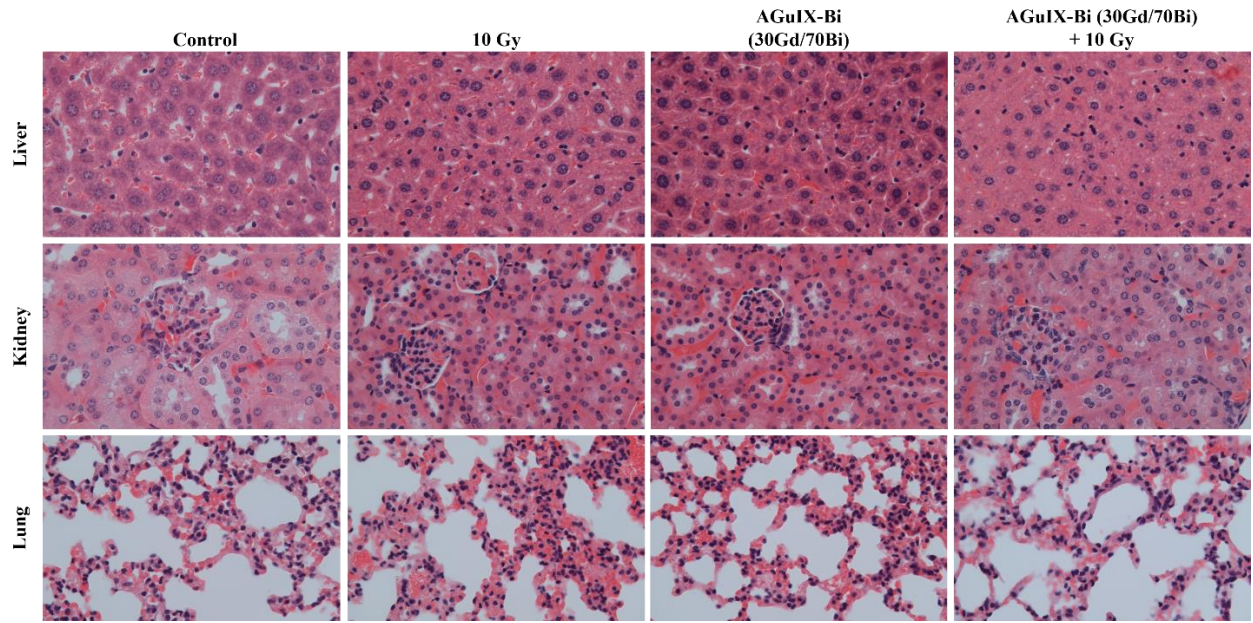

**Figure S24:** Histopathology of mice treated with either saline or AGuIX-Bi (30Gd/70Bi) and/or 10 Gy radiation 1 mo post-treatment. No long-term toxicity was identified in lungs, kidney, or liver.

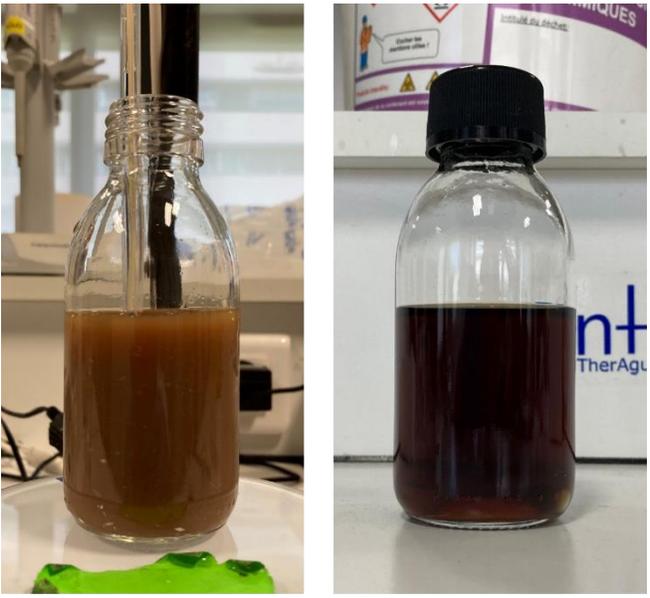

    **A**        **B**

**Figure S25:** A) Image of the solution after addition of the BiCl$_3$. At high concentrations of BiCl$_3$, the solution precipitates. B) Clear solution after pH adjustment and stirring for 48 hours at 80°C.

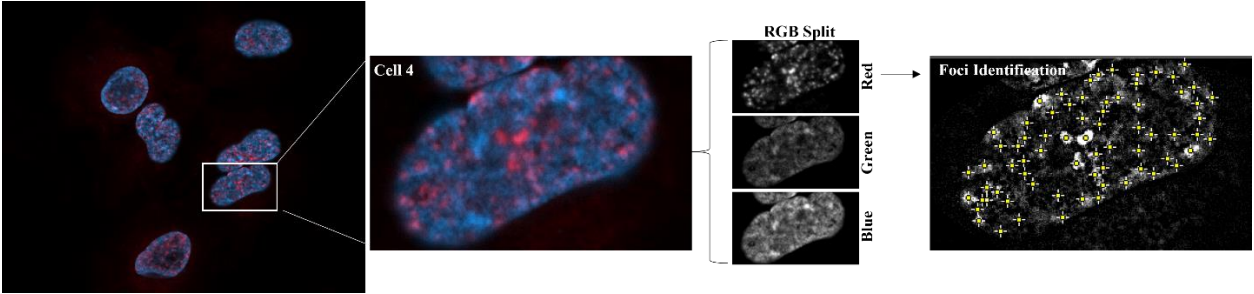

**Figure S26:** Representative example of ImageJ automated macro steps for γH2AX foci quantification.

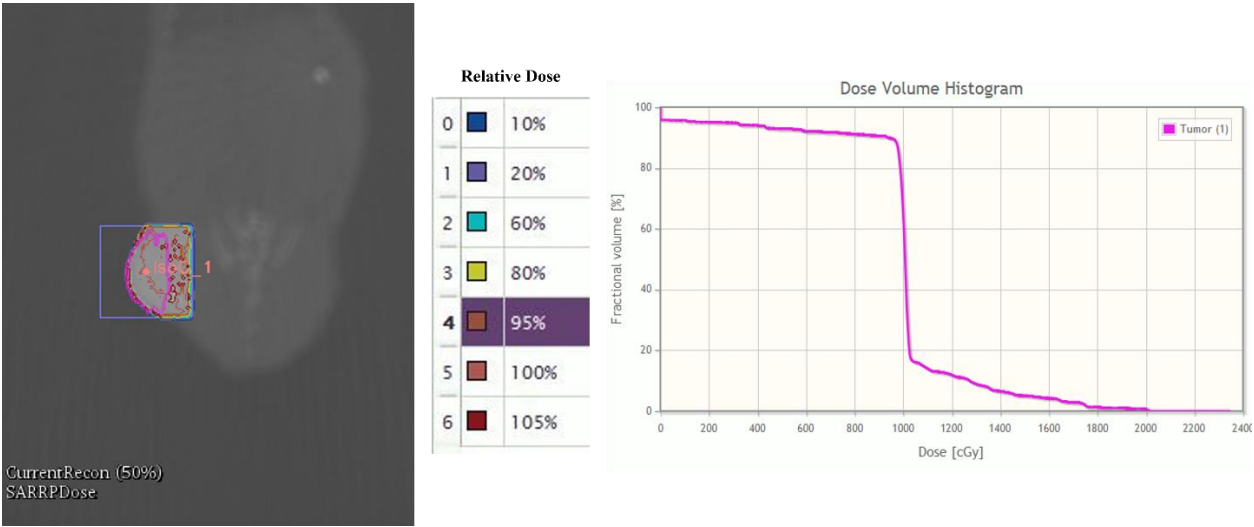

**Figure S27:** Representative SARRP treatment plan for a mouse given 10 Gy radiation. A cone-beam computed tomography image was acquired at 60 kVp (0.8 mA current with a 1 mm Al filter) to identify tumor isocenter in MuriSlice followed by a single posterior-anterior 220 kVp (13 mA current with a 0.15 mm Cu filter) beam delivering a dose of 10 Gy to a 1 x 1 cm$^2$ collimated field.

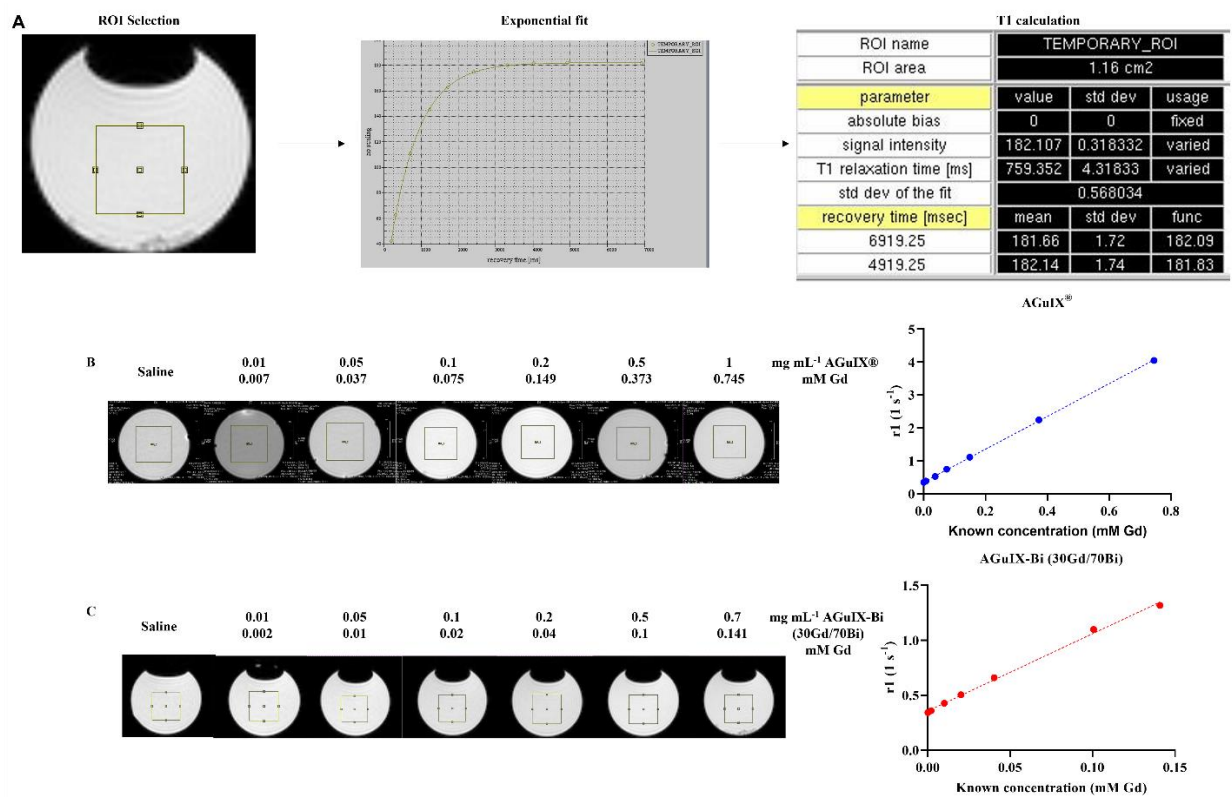

**Figure S28:** A) Representative example of a phantom of AGuIX-Bi (30Gd/70Bi) used to determine nanoparticle relaxivity at 7 T. Bruker Paravision 6.0.1 was used for MRI data acquisition and T1 map data analysis to determine $r_1$ for B) AGuIX® and C) AGuIX-Bi (30Gd/70Bi).